\RequirePackage{fix-cm}
\documentclass[twocolumn]{svjour3}  
\smartqed

\usepackage{graphicx}
\usepackage{color}
\usepackage{amsmath}
\usepackage{amsfonts}

 \RequirePackage{mathptmx} 
 \RequirePackage{latexsym}
 \RequirePackage{amsmath}
 \RequirePackage{amssymb}
 \RequirePackage[numbers,sort&compress]{natbib}
\RequirePackage[colorlinks,citecolor=blue,urlcolor=blue,linkcolor=blue]{hyperref}

\usepackage[colorlinks,citecolor=blue,urlcolor=blue,linkcolor=blue]{hyperref}
\usepackage[numbers,sort&compress]{natbib}

\newcommand*{\doi}[1]{\href{http://dx.doi.org/#1}{doi: #1}}

\usepackage{subfig}

\usepackage{hyphenat}

\begin{document}

\title{Charge superradiance on charged BTZ black holes}

\author{Sebastian Konewko         \and
        Elizabeth Winstanley 
}

\institute{S.~Konewko \at Department of Applied Physics, Gent University, Sint Pietersnieuwstraat 41, 9000 Gent, Belgium \\ \email{sebastian.konewko@ugent.be} \and E.~Winstanley \at Consortium for Fundamental Physics, School of Mathematics and Statistics,  The University of Sheffield, Hicks Building, Hounsfield Road, Sheffield. S3 7RH United Kingdom \\ \email{E.Winstanley@sheffield.ac.uk}
\\	
}

\date{}

\maketitle 

\begin{abstract}
We study superradiant scattering for a charged scalar field  subject to Robin (mixed) boundary conditions on a charged BTZ black hole background. 
Scalar field modes having a real frequency do not exhibit superradiant scattering, independent of the boundary conditions applied. 
For scalar field modes with a complex frequency, no superradiant scattering occurs if the black hole is static.
After exploring some regions of the parameter space, we provide evidence for the existence of superradiantly scattered modes with complex frequencies for a charged and rotating BTZ black hole. 
Most of the superradiantly scattered modes we find satisfy Robin (mixed) boundary conditions, but there are also superradiantly scattered modes with complex frequencies satisfying Dirichlet and Neumann boundary conditions.
We explore the effect of the black hole and scalar field charge on the outgoing energy flux of these superradiantly scattered modes, and also investigate their stability.
\end{abstract}

\section{Introduction}
\label{sec:intro}

In the scattering of waves incident on a black hole, superradiant scattering occurs if the reflected wave has greater amplitude than the incident wave \cite{Brito:2015oca}. 
For example, low-frequency bosonic waves interacting with a rotating Kerr black hole exhibit superradiant scattering \cite{Misner:1972kx,Press:1972zz,Chandrasekhar:1985kt,Teukolsky:1974yv}.
This phenomenon is the wave analogue of the Penrose process for particles \cite{Penrose:1971uk}, with the wave extracting rotational energy from the black hole. 
A similar effect occurs for charged scalar field waves scattered by a charged Reissner-Nordstr\"om black hole \cite{Bekenstein:1973mi,Nakamura:1976nc,DiMenza:2014vpa,Benone:2015bst,DiMenza:2019zli}.
In this case it is electrostatic rather than rotational energy which is extracted from the black hole by the wave. 
The amplification of superradiantly scattered scalar field modes on a Kerr black hole is comparatively small, for example for a Kerr black hole with angular momentum parameter $a=0.99M$ (where $M$ is the black hole mass), the maximum amplification factor for a massless scalar field mode is about $0.4$\% \cite{Brito:2015oca}. 
In contrast, charged scalar field modes undergoing superradiant scattering on a Reissner--Nordstr\"om black hole can achieve significantly larger amplification factors, up to $40$\% \cite{Brito:2015oca}. 
For this reason, charged superradiant scattering is a very useful toy model for numerical studies. The fact that the Reissner-Nordstr\"om black hole is static and spherically symmetric also simplifies the mathematical analysis of charged superradiant scattering compared with that on rotating black holes.

One consequence of superradiant scattering is the possibility that it can lead to a superradiant instability, if there is some mechanism  confining a superradiantly amplified field \cite{Press:1972zz}. 
For example, a massive scalar field on a rotating Kerr black hole has a potential which is confining and exhibits a superradiant instability \cite{Detweiler:1980uk,Zouros:1979iw,Dolan:2007mj}.
However, superradiantly unstable scalar field modes on a Kerr black hole grow only very slowly in time, which presents major challenges for numerical studies of the evolution of the superradiant instability \cite{Okawa:2014nda,East:2013mfa}.

For a charged scalar field on a Reissner-Nordstr\"om black hole, the presence of a mass is not sufficient to confine the field and cause an instability \cite{Hod:2013nn,Hod:2015hza}.
An alternative confining mechanism is therefore required if there is to be a superradiant instability.
In the ``black hole bomb'' scenario \cite{Press:1972zz}, the confining mechanism is provided by a perfectly reflecting mirror. 
For a charged scalar field on a Reissner-Nordstr\"om black hole, there is an instability if the mirror is sufficiently far from the event horizon of the black hole \cite{Herdeiro:2013pia,Degollado:2013bha}. 
This instability can grow many times faster than the corresponding instability of a massive scalar field on a Kerr black hole, which enables the numerical evolution of the instability, leading to the formation of a black hole with scalar hair \cite{Sanchis-Gual:2015lje,Sanchis-Gual:2016tcm}.

An alternative confining mechanism can arise when considering asymptotically anti-de Sitter (adS) rather than asym-ptotically flat black holes.
In this situation superradiant scattering persists, and the adS boundary can act like a confining mirror. 
A massless scalar field on a sufficiently small rotating asymptotically-adS black hole is superradiantly unstable, and the same is true for a charged scalar field on a sufficiently small Reissner-Nordstr\"om-adS black hole (see, for example, \cite{Cardoso:2004hs,Cardoso:2006wa,Kunduri:2006qa,Kodama:2007sf,Gubser:2008px,Aliev:2008yk,Murata:2008xr,Kodama:2009rq,Uchikata:2009zz,Dias:2011tj,Li:2012rx,Cardoso:2013pza,Wang:2014eha,Wang:2015fgp,Delice:2015zga,Green:2015kur,Bosch:2016vcp,Dias:2016pma,BarraganAmado:2018zpa,Chesler:2018txn,Li:2019tns,Katagiri:2020mvm,Chesler:2021ehz,Ishii:2021ncf,Ishii:2022lwc} for a selection of papers from the extensive literature on this topic).  

The three-dimensional, asymptotically adS, BTZ black hole \cite{Banados:1992gq,Banados:1992wn,Carlip:1995qv} has been widely studied as a toy model for phenomena on black holes in four or more space-time dimensions. 
In particular, one can consider whether superradiant scattering, or a superradiant instability, is possible on a BTZ black hole. 
The analysis of these questions for a neutral scalar field is greatly simplified by the fact that the field modes can be written in terms of hypergeometric functions (see, for example, \cite{Ichinose:1994rg,Ghoroku:1994np}).
Since null infinity is a time-like hypersurface in adS, boundary conditions have to be applied to a perturbing field. 
For a neutral scalar field, applying the simplest boundary conditions, namely Dirichlet boundary conditions, superradiant scattering is absent \cite{Ortiz:2011wd}.
Superradiant scattering also does not occur for a fermion field vanishing on the boundary \cite{Ortiz:2018ddt}.
However, one can impose more general (Neumann or Robin) boundary conditions on the scalar field \cite{Ishibashi:2004wx}, and, for at least some Robin boundary conditions, there exist modes which exhibit superradiant scattering \cite{Dappiaggi:2017pbe}. 
The impact of Robin boundary conditions on superradiant scattering has also been considered on four-dimensional Reissner-Nordstr\"om-adS \cite{Katagiri:2020mvm} and  Kerr-adS black holes \cite{Ferreira:2017tnc}. 

In this paper we examine whether superradiant scattering exists for a charged scalar field subject to Robin (mixed) boundary conditions on a charged analogue of the BTZ black hole.
We seek to understand whether the analogy between superradiant scattering behaviour on a BTZ black hole and four dimensional black holes persists for a charged scalar field.
In particular, we ascertain whether there is charged superradiant scattering on a static charged BTZ black hole. We also examine whether superradiant amplification is enhanced in this set-up compared with superradiant scattering on a rotating BTZ black hole.

We begin, in Section~\ref{sec:scalar},  by reviewing the black hole metric and separable solutions of the charged scalar field equation on this spacetime background, paying particular attention to the boundary conditions satisfied by the field far from the black hole.
In Section~\ref{sec:superradiance} we study the possibility of superradiant scattering using two approaches: firstly a Wronskian condition (arising from the conservation of the charge current) which is valid for waves having real frequency, and secondly, considering the fluxes of energy and charge down the horizon for an ingoing wave.
In particular, an ingoing wave which is superradiantly scattered will have an outgoing energy flux. 
Using the Wronskian, we find that there is no superradiant scattering for field modes having real frequency.
If the frequency is complex, by considering the energy flux, we show that superradiant scattering is absent if the black hole is nonrotating.
This leaves open the possibility of superradiant scattering for charged, rotating black holes, which is studied in Section~\ref{sec:rotating}.
Using a simple numerical method, valid for frequencies in the superradiant regime, we provide evidence for the existence of charged scalar field modes which are superradiantly scattered. 
We explore the effect of increasing either the black hole or scalar field charge on the energy fluxes of these superradiantly scattered modes, as well as investigating their stability.
Our conclusions are presented in Section~\ref{sec:conc}.
Two appendices give further details of our numerical procedure and the complex frequencies of the superradiantly scattered modes.

\section{Charged scalar field on a charged BTZ black hole}
\label{sec:scalar}

\subsection{Charged BTZ black holes}
\label{sec:BTZ}

The neutral BTZ black hole \cite{Banados:1992gq,Banados:1992wn,Carlip:1995qv} is a solution of the three-dimensional Einstein equations with a negative cosmological constant $\Lambda = -\ell ^{-2}$, having metric
\begin{subequations}
	\label{eq:BTZneutral}
\begin{align}
	ds^{2}  = & ~ -N_{0}(r) \, dt^{2} + N_{0}(r)^{-1} \, dr^{2} 
 + r^{2} \left[ d\varphi  + N_{0}^{\varphi}(r) \, dt \right]^{2}
	\label{eq:BTZneutralmetric}
\end{align}
where
\begin{equation}
 N_{0} (r) = \frac{r^{2}}{\ell ^{2}} - M + \frac{J^{2}}{4r^{2}}, \qquad
 N_{0}^{\varphi } (r) = -\frac{J}{2r^{2}} ,
 \label{eq:N0}
\end{equation}
\end{subequations}
with $M$ the mass and $J$ the angular momentum of the black hole. 

In the static case ($J=0$), the black hole acquires an electric charge $Q$ by introducing the electromagnetic potential
$A_{\mu }=A_{0}\delta _{\mu }^{t}$, where
\begin{subequations}
	\label{eq:staticcharged}
\begin{equation}
	A_{0} = -Q \ln \left( \frac {r}{r_{0}} \right) ,
	\label{eq:staticgaugepot}
\end{equation}
and an arbitrary length scale $r_{0}$ has been introduced to render the argument of the logarithm dimensionless. The metric for a static charged BTZ black hole is then 
\cite{Banados:1992gq,Banados:1992wn,Carlip:1995qv}
\begin{equation}
	ds^{2} = -N(r) \, dt^{2} + N(r)^{-1} \, dr^{2} + r^{2} \, d\varphi ^{2}
	\label{eq:staticchargedmetric}
\end{equation}
and the lapse function takes the form
\begin{equation}
	N(r) = \frac{r^{2}}{\ell ^{2}} - M - \frac{Q^{2}}{2}\ln \left( \frac{r}{r_{0}} \right)  .
	\label{eq:chargedlapse}
\end{equation}
\end{subequations}

The generalization of the rotating BTZ black hole to include an electric charge is far from straightforward \cite{Carlip:1995qv,Garcia:1999py}. Various three-dimensional, charged, rotating black holes are presented in \cite{Clement:1993kc,Clement:1995zt,Kamata:1995zu,Cataldo:1999fh,Martinez:1999qi}.
In this paper we consider the following charged generalization of the BTZ metric \cite{Martinez:1999qi}:
\begin{subequations}
	\label{eq:rotatingcharged}
	\begin{align}
		ds^{2}  =  & ~ 
		- N(r) \frac{r^{2}}{R(r)^{2}} \, dt^{2} + N(r)^{-1}\, dr^{2} 
		\nonumber \\ & ~ 
  + R(r)^{2} \left[ d\varphi + N^{\varphi }(r) \, dt \right] ^{2}
		\label{eq:rotatingchargedmetric}
	\end{align}
where $N(r)$ is the same as in the static case (\ref{eq:chargedlapse}) and the other functions appearing in the metric are:
\begin{align}
	R(r)^{2} = & ~
	r^{2} + \frac{\Omega ^{2}\ell ^{2}}{1- \Omega ^{2}} \left[ M + \frac{Q^{2}}{2}\ln \left( \frac{r}{r_{0}} \right)  \right] ,
	\nonumber
	\\
	N^{\varphi }(r) = & ~
	-\frac{\Omega \ell }{\left( 1- \Omega ^{2} \right) R(r)^{2}} \left[ M + \frac{Q^{2}}{2}\ln \left( \frac{r}{r_{0}} \right)  \right] ,
	\label{eq:rotatingchargedfunctions}
\end{align}
\end{subequations}
where $M$, $Q$ and $\Omega \in [0,1)$ are constants.
The mass ${\widetilde {M}}$ and angular momentum ${\widetilde {J}}$ of the black hole  are given in terms of the parameters $M$, $Q$, and $\Omega $ as follows \cite{Martinez:1999qi}:
\begin{align}
{\widetilde {M}} = & ~ \frac{1}{1-\Omega ^{2}} \left[ M \left(1 + \Omega ^{2} \right) - \frac{1}{2} Q^{2}\Omega ^{2} \right] ,
\nonumber \\
{\widetilde {J}} = & ~ \frac{2\Omega \ell }{1-\Omega ^{2}} \left[ M - \frac{1}{4}Q^{2} \right] .
\label{eq:MJ}
\end{align}
Unlike the neutral BTZ black hole, the metric (\ref{eq:rotatingchargedmetric}) cannot be obtained by identifying points in three-dimensional adS space-time. In particular, the scalar curvature ${\mathcal {R}}$ is not constant:
\begin{equation}
	{\mathcal {R}} =  \frac{Q^{2}}{2r^{2}} - \frac{6}{\ell ^{2}}.
	\label{eq:ricciscalar}
\end{equation}

When $\Omega >0$, the electromagnetic potential $A_{\mu }$ acquires a nonzero magnetic part and is given by \cite{Martinez:1999qi}:
\begin{equation}
	A_{\mu } dx^{\mu } = - \frac{Q}{{\sqrt {1-\Omega ^{2}}}} \left[ dt - \Omega \ell \, d\varphi \right] \ln \left( \frac{r}{r_{0}} \right) .
	\label{eq:rotatinggaugepot}
\end{equation}
The nonzero components of the electromagnetic field strength tensor $F_{\mu \nu }=\partial _{\mu }A_{\nu } - \partial _{\nu }A_{\mu }$ are then
\begin{equation}
    F_{rt} =- \frac{Q}{r{\sqrt {1-\Omega ^{2}}}} , 
   \qquad 
    F_{r\varphi }  = \frac{Q\Omega \ell }{r{\sqrt {1-\Omega ^{2}}}} .
    \label{eq:Faraday}
\end{equation}
For comparison, the electrostatic potential for a four-dimens-ional Reissner-Nordstr\"om black hole with charge $Q_{\text{RN}}$ is, for a suitable choice of gauge, 
\begin{equation}
    A_{\mu }dx^{\mu } = -\frac{Q_{\text{RN}}}{r} \, dt,
\end{equation}
from which we find
\begin{equation}
    F_{rt} = \frac{Q_{\text{RN}}}{r^{2}},
\label{eq:RNE}
\end{equation}
giving the usual Coulomb law for the electric field due to a charge $Q_{\text{RN}}$ located at the origin in three spatial dimensions. 
Therefore $Q_{\text{RN}}$ is  the charge of the Reissner-Nordstr\"om black hole.
Comparing (\ref{eq:Faraday}, \ref{eq:RNE}), it can be seen that there is a sign difference in $F_{rt}$. 
The electric field component $F_{rt}$ in (\ref{eq:Faraday}) has the expected $r^{-1}$ behaviour for a point charge at the origin in two spatial dimensions, but is proportional to $-Q$ rather than $Q$. 
As a result, the charge ${\widetilde {Q}}$ of the black hole is given by 
\begin{equation}
{\widetilde {Q}} = - \frac{Q}{{\sqrt {1 - \Omega ^{2}}}} .
    \label{eq:charge}
\end{equation}
The fact that the black hole charge ${\widetilde {Q}}$ has the opposite sign to $Q$ will have important consequences for the interpretation of the charged scalar field modes which we study later in this paper.

In the limit $\Omega \rightarrow 0$, the metric (\ref{eq:rotatingcharged}) and gauge field potential (\ref{eq:rotatinggaugepot}) reduce to those in the static case (\ref{eq:staticcharged}). 
If we set $Q=0$, the metric (\ref{eq:rotatingchargedmetric}) does not reduce to the original rotating BTZ metric (\ref{eq:BTZneutralmetric}) in $(t,r,\varphi )$ coordinates.  
However, using $R$ as the radial coordinate, when $Q=0$ the metric (\ref{eq:rotatingchargedmetric}) becomes
\begin{subequations}
\begin{align}
	ds^{2} = & ~ - {\widetilde {N}}(R) \, dt^{2} + {\widetilde {N}}(R)^{-1} \, dR^{2}
	+ R^{2} \left[ d\varphi + N^{\varphi }(r) \, dt \right] ^{2}
\end{align}
where we have defined the function
\begin{align}
	{\widetilde {N}}(R) =  & ~ N(r) \frac{r^{2}}{R^{2}} 
= \frac{R^{2}}{\ell ^{2}} - \frac{M\left(1 + \Omega ^{2}\right) }{1-\Omega ^2} 
	+ \frac{M^{2}\Omega ^{2}\ell ^{2}}{\left(1 - \Omega ^{2} \right) ^{2}R^{2}} .
 \end{align}
\end{subequations}
We therefore have a metric of the form (\ref{eq:BTZneutralmetric}) with mass  ${\widetilde {M}}= M\left(1 + \Omega ^{2}\right)/\left(1 -\Omega ^{2} \right)^{2} $ and angular momentum ${\widetilde {J}}=$ \\ $2M\Omega \ell/\left(1 - \Omega ^{2} \right)$, in accordance with (\ref{eq:MJ}).

The horizons of the black hole are located at those values of the radial coordinate $r$ for which $N(r)$ vanishes. If $M< Q^{2}\left[ 1-2\ln \left( Q\ell /2r_{0} \right) \ \right] /4 $ there is a naked singularity at $r=0$; we do not consider this possibility further. For $M>Q^{2}\left[ 1-2\ln \left( Q\ell /2r_{0} \right) \right] /4$ there is an event horizon at $r=r_{h}$, the largest zero of $N(r)$ and an inner horizon at the smaller positive zero of $N(r)$. These two horizons coincide when $M=Q^{2}\left[ 1-2\ln \left( Q\ell /2r_{0} \right) \right] /4$ and in this case we have an extremal black hole \cite{Tang:2016vmu}.  
In this paper we focus on the case where the black hole is nonextremal.

By making a gauge transformation of the form
\begin{align}
	A_{\mu } \rightarrow  & ~ A_{\mu } + \partial _{\mu } \chi, 
 \nonumber \\ 
	\chi = & ~ \frac{Q}{{\sqrt {1-\Omega ^{2}}}} \left(t - \Omega \ell \varphi \right)  
	\ln \left(  \frac{r_{h}}{r_{0}}\right),
 \label{eq:gaugetransformation}
\end{align}
we may set $r_{0}=r_{h}$ without loss of generality. We then find, by considering the zeros of (\ref{eq:chargedlapse}),  that
\begin{equation}
M= \frac{r_{h}^{2}}{\ell ^{2}}.
\label{eq:M}
\end{equation}
At the horizon, we have $R(r_{h})^{2} = r_{h}^{2}/\left( 1-\Omega ^{2}\right) $ and
$N^{\varphi }(r_{h}) = -\Omega /\ell $, so that $\Omega /\ell $ is the angular speed with which the event horizon rotates.  

In our analysis of superradiant scattering, we will be interested in the flux of energy down the event horizon of the black hole. For this analysis, we will require suitable coordinates which are regular across the horizon. 
We will employ ingoing Eddington-Finkelstein (EF) coordinates. First we define an ingoing null coordinate $v$ by
\begin{subequations}
	\label{eq:EFcoordinates}
\begin{equation}
 dv = dt+\frac{1}{r} \frac{R(r)}{N(r)} \, dr ,
\end{equation}
and a new angular coordinate ${\widehat{\varphi}}_{v}$ by
\begin{equation}
	d{\widehat{\varphi}}_{v}= d\varphi - \frac{R(r)}{r} \frac{N^{\varphi }(r)}{N(r)} \, dr.
\end{equation}
\end{subequations}
Then the coordinates 
$(v,r,{\widehat{\varphi}}_{v})$ are ingoing EF coordinates, in terms of which the metric (\ref{eq:rotatingchargedmetric}) becomes
\begin{align}
\label{eq:EFmetric}
	ds^{2}= & ~ -N(r)\frac{r^2}{R(r)^{2}}\, dv^{2} + \frac{2r}{R(r)} \, dv \, dr 
	\nonumber \\ & ~
 + R(r)^{2} \left[ d{\widehat{\varphi}}_{v} + N^{\varphi }(r) \, dv  \right] ^{2}.
\end{align}
The metric (\ref{eq:EFmetric}) is regular when $r=r_{h}$ and $N(r)=0$, as required.
Near the horizon, the ingoing  EF coordinates take the form
\begin{equation}
\label{eq:EFtransform}
 v = t+\frac{r_{*}}{\sqrt {1-\Omega ^{2}}}, \qquad 
 {\widehat{\varphi}}_{v} = \varphi +  \frac{\Omega r_{*}}{\ell {\sqrt {1-\Omega ^{2}}}},
\end{equation}
where $r_{*}$ is the usual tortoise coordinate, defined by 
\begin{equation}
	\frac{dr_{*}}{dr} = \frac{1}{N(r)}.
	\label{eq:tortoise}
\end{equation}
Similarly, we can define outgoing EF coordinates $(u,r,{\widehat {\varphi }}_{u})$ as follows:
\begin{align}
 du &  = dt-\frac{1}{r} \frac{R(r)}{N(r)} \, dr ,
 \nonumber \\ 
 d{\widehat{\varphi}}_{u} & = d\varphi +\frac{R(r)}{r} \frac{N^{\varphi }(r)}{N(r)} \, dr.
 \end{align}
 Near the horizon, the outgoing EF coordinates take the form
 \begin{equation}
   u = t-\frac{r_{*}}{\sqrt {1-\Omega ^{2}}},  
   \qquad 
   {\widehat{\varphi}}_{u} = \varphi -  \frac{\Omega r_{*}}{\ell {\sqrt {1-\Omega ^{2}}}}.
   \label{eq:EFtransform1}
\end{equation}

\subsection{Charged scalar field}
\label{sec:chargeS}

We consider a scalar field $\Phi $ with charge $q$ and mass $m$ propagating on the rotating charged black hole (\ref{eq:rotatingchargedmetric}), and satisfying the charged scalar field equation
\begin{equation}
	\left[ D_{\mu }D^{\mu } - m^{2}  \right] \Phi =0,
	\label{eq:scalareqn}
\end{equation}
where $D_{\mu }=\nabla _{\mu }-iqA_{\mu }$ is the covariant derivative.
We assume that the scalar field is minimally coupled to the geometry.
The stress-energy tensor for the charged scalar field is
\begin{align}
	T_{\mu \nu } = & ~
	\Re \left\{ 
	\left( D_{\mu } \Phi \right) ^{*} D_{\nu }\Phi 
- \frac{1}{2} g_{\mu \nu }g^{\rho \sigma }
	\left(  D_{\rho } \Phi \right) ^{*}D_{\sigma }\Phi 
	\right. \nonumber \\ & ~ \left.
	-\frac{1}{2}m^{2}g_{\mu \nu }\Phi ^{*}\Phi 
	\right\} ,
	\label{eq:SET}
\end{align}
where $\Re $ is the real part and a star is used to denote complex conjugation.  
The scalar field has an associated current density $J^{\mu }$ given by 
\begin{equation}
    J^{\mu } = -iq \left[ \Phi ^{*} \left( D^{\mu } \Phi \right)  - \Phi \left( D^{\mu } \Phi \right) ^{*}\right] .
    \label{eq:current}
\end{equation}
In this paper we treat the scalar field as a test field. However, if we were to go beyond this approximation, and consider the effect of the charged scalar field on the electromagnetic field, the current density $J^{\mu }$ (\ref{eq:current}) would act as source term in Maxwell's equations:
\begin{equation}
    \nabla _{\mu }F^{\mu \nu } = -4\pi J^{\nu }.
\end{equation} 
In Section~\ref{sec:superradiance} we shall study in particular the radial component $J^{r}$ of the current density (\ref{eq:current}).

Mode solutions of the scalar field equation (\ref{eq:scalareqn}) take the form
\begin{equation}
	\Phi_{\omega k} (t, r, \varphi ) = \frac{1}{{\sqrt {r}}} \ e^{-i\omega t} \ e^{ik\varphi } \ X_{\omega k} (r) ,
	\label{eq:mode} 
\end{equation}
where $\omega $ is the frequency of the wave (which may be complex) and $k\in {\mathbb {Z}}$ is the azimuthal quantum number. 
In terms of the tortoise coordinate $r_{*}$ (\ref{eq:tortoise}),
the radial function $X_{\omega k} (r)$ satisfies the equation
\begin{subequations}
	\label{eq:radial}
\begin{equation}
	\left[ \frac{d^{2}}{dr^{2}_{*}} + V_{\omega k}(r) \right] X_{\omega k} (r) = 0
	\label{eq:radialeq}
\end{equation}
where the potential $V_{\omega k}(r)$ takes the form
\begin{align}
	V_{\omega k} (r) = & ~
	\left[  
	\frac{\omega - k\Omega \ell ^{-1}}{\sqrt {1-\Omega ^{2}}}  - qQ \ln \left( \frac{r}{r_{h}} \right) 
	\right] ^{2}
 \nonumber \\ & ~
 - m^{2} N(r)
	+ \frac{N(r)^{2}}{4r^{2}}
	- \frac{N'(r)N(r)}{2r}
 \nonumber \\ & ~
	- \frac{\left(\omega \Omega -k\ell ^{-1} \right) ^{2} \ell ^{2} N(r)}{\left( 1 - \Omega ^{2} \right) r^{2}} .
	\label{eq:potential}
\end{align}
\end{subequations}
As $r\rightarrow r_{h}$ and the event horizon is approached, we have $r_{*} \rightarrow -\infty $ and 
\begin{equation}
	V_{\omega k}(r) \rightarrow {\widetilde {\omega }}^{2} 
	\label{eq:potentialhorizon}
\end{equation}
where we have defined
\begin{equation}
	{\widetilde {\omega }} = \frac{\omega - k\Omega \ell ^{-1}}{\sqrt {1-\Omega ^{2}}} .
	\label{eq:omegatilde} 
\end{equation}
Therefore, near the horizon, the radial function $X_{\omega k}(r)$ takes the form
\begin{equation}
	X_{\omega k}(r) \sim A_{\omega k}e^{i{\widetilde {\omega }}r_{*}} + B_{\omega k}e^{-i{\widetilde {\omega }}r_{*}}
	\label{eq:Xhorizon}
\end{equation}
where $A_{\omega k}$ and $B_{\omega k}$ are complex constants.
The frequency of the wave has effectively been shifted due to the rotation of the black hole. The fact that ${\widetilde{\omega }}$ does not depend on the charge stems from our choice of gauge, in that the electromagnetic gauge potential (\ref{eq:rotatinggaugepot}) vanishes at the horizon since we have taken $r_{0}=r_{h}$.

Far from the black hole, as $r\rightarrow \infty $, the leading-order behaviour of the potential (\ref{eq:potential}) is, in general,
\begin{equation}
	V_{\omega k} (r) \sim - \left[ m^{2}  + \frac{3}{4 \ell^{2}} \right] \frac{r^{2}}{\ell ^{2}}.
	\label{eq:potentialinfinitygen}
\end{equation}
This leading-order behaviour is the same as for the neutral scalar field, and does not depend on the frequency $\omega $ or the azimuthal quantum number $k$.
In this regime the tortoise coordinate has the following form:
\begin{equation}
	r_{*} \sim - \frac{\ell ^{2}}{r},
\end{equation}
yielding the equidimensional differential equation
\begin{equation}
	\left[ \frac{d^{2}}{dr^{2}_{*}} - \frac{\mu ^{2}}{r_{*}^{2}}  \right] X_{\omega k} (r) = 0
	\label{eq:radialinfinity}
\end{equation}
where $\mu ^{2}$ is a constant given by
\begin{equation}
	\mu ^{2} =  m^{2}\ell ^{2}  +\frac{3}{4} .
\end{equation}
Let us assume for the moment that $\mu ^{2}\neq 0$. The solutions of (\ref{eq:radialinfinity}) are $X_{\omega k}\sim r_{*}^{p}\sim r^{-p}$, where
\begin{equation}
	p = \frac{1}{2} \left( 1 \pm {\sqrt{ 1+4\mu ^{2}}}  \right) .
	\label{eq:p}
\end{equation}
For $4\mu ^{2}>-1$, the values of $p$ are real and 
\begin{align}
	X_{\omega k} (r) 
	\sim  & ~ C_{\omega k} r^{-\frac{1}{2}(1+{\sqrt {1+4\mu ^{2}}})}
	+ D_{\omega k }r^{-\frac{1}{2}(1-{\sqrt {1+4\mu ^{2}}})}
	\label{eq:Xinfinity1}
\end{align}
for complex constants $C_{\omega k}$, $D_{\omega k}$.
The second term gives a scalar field mode \eqref{eq:mode} which diverges at infinity when $4\mu ^{2}>3$ and therefore we set $D_{\omega k}=0$ in this case. 
In this situation there is no choice of boundary conditions which can be imposed on the scalar field at infinity. 

For $-1<4\mu^{2}<3$, both solutions in (\ref{eq:Xinfinity1}) give regular scalar field modes \eqref{eq:mode}, resulting in some freedom in the choice of boundary conditions at infinity \cite{Bussola:2017wki,Garbarz:2017wzv}.
The solution with $D_{\omega k}=0$ satisfies Dirichlet boundary conditions, 
while, following \cite{Bussola:2017wki}, we define Neumann boundary conditions to be such that  $C_{\omega k}=0$. 
If both $C_{\omega  \ell }$ and $D_{\omega \ell }$ are nonzero, then we have Robin (mixed) boundary conditions. 
In this situation we write the solution (\ref{eq:Xinfinity1}) in the form \cite{Dappiaggi:2017pbe,Bussola:2017wki}
\begin{align}
	X_{\omega k} (r)
	\sim  & ~ E_{\omega k} \left[ r^{-\frac{1}{2}(1+{\sqrt {1+4\mu ^{2}}})} \cos \zeta
	+ r^{-\frac{1}{2}(1-{\sqrt {1+4\mu ^{2}}})} \sin \zeta \right] 
	\label{eq:XinfinityRobin} 
\end{align}
where $E_{\omega k}$ is a complex constant and the real angle $\zeta $  (which we term the ``Robin parameter'')  can be taken to lie in the interval $0\le \zeta < \pi $ (we could equally well take $\zeta \in (-\frac{\pi }{2}, \frac{\pi }{2}]$).
Setting $\zeta = 0$ yields Dirichlet boundary conditions, while $\zeta  = \frac{\pi }{2}$ corresponds to Neumann boundary conditions. 

When $4\mu ^{2}=-1$, we have
\begin{equation}
	X_{\omega k}(r) \sim C_{\omega k}r^{-\frac{1}{2}} + D_{\omega k}r^{-\frac{1}{2}} \ln \left( \frac{r}{r_{h}}\right) .
	\label{eq:Xinfinity2}
\end{equation}
Both solutions are square-integrable in this case, so again we have a choice of boundary conditions.
For $4\mu ^{2}<-1$, the exponent $p$ (\ref{eq:p}) is complex and $X_{\omega k}(r)$ is oscillatory. Once again both linearly independent solutions of the radial equation are square-integrable at infinity. 
However, these values of $\mu^{2}$ violate the Breitenlohner-Freedman bound \cite{Breitenlohner:1982bm,Breitenlohner:1982jf} and therefore we do not consider them further in this work.

The above discussion of the boundary conditions at infinity is valid only when $\mu ^{2}\neq 0$, in which case the behaviour of the charged scalar field at infinity is identical to that for a neutral scalar field.  In the special case $\mu ^{2}=0$ the leading order behaviour of the potential (\ref{eq:potential}) is no longer (\ref{eq:potentialinfinitygen}), but instead we have, as $r\rightarrow \infty$, 
\begin{equation}
	V_{\omega k}(r) \sim q^{2}Q^{2} \left[ \ln \left( \frac{r}{r_{h}}\right)\right] ^{2}. 
	\label{eq:potentialinfinitymuzero}
\end{equation}
In this case it is not possible to solve the asymptotic form of the radial equation exactly in terms of elementary functions. However, it is possible to perform an asymptotic expansion for the radial function $X_{\omega k}(r)$ in this case. The first couple of terms in this asymptotic expansion are:
\begin{align}
	X_{\omega \ell }(r) \sim  & ~ C_{\omega k} \left\{ \frac{1}{r} - \frac{q^{2}Q^{2}\ell ^{4}}{6r^{3}} \left[ \ln \left( \frac{r}{r_{h}} \right) \right] ^{2} +\ldots \right\}
	\nonumber \\ & ~
	\hspace{-0.5cm} +  D_{\omega k} \left\{ 1 - \frac{q^{2}Q^{2}\ell ^{4}}{2r^{2}} \left[ \ln \left( \frac{r}{r_{h}} \right) \right] ^{2} +\ldots \right\} .
	\label{eq:Xinfinity3}
\end{align} 
The second solution gives a mode which is not square integrable at infinity, so we set $D_{\omega k}=0$ in this case. 
The behaviour at infinity of a massless and conformally coupled charged scalar field is thus rather different from that seen in the neutral case. 

\section{Criterion for superradiant scattering}
\label{sec:superradiance}

We now explore whether superradiant scattering occurs for a charged scalar field, examining separately the cases where the frequency $\omega $ is real or complex.

\subsection{Real frequency}
\label{sec:Wronskian}

We first consider the situation in which the frequency $\omega $ is real and follow an approach of \cite{Katagiri:2020mvm, Ortiz:2011wd}.
Near the horizon of an eternal charged BTZ black hole, a scalar field wave will have a radial function of the form (\ref{eq:Xhorizon}), giving
\begin{align}
    \Phi _{\omega k} & \sim \frac{1}{{\sqrt {r}}} \left\{ 
    A_{\omega k }\exp  \left[  -i(\omega t - {\widetilde {\omega }}r_{*}- k\varphi ) \right]
    \right. \nonumber \\ & \left.  \qquad 
    + B_{\omega k} \exp \left[ -i(\omega t +{\widetilde {\omega }}r_{*} - k\varphi )  \right]
    \right\} .
\end{align}
Using (\ref{eq:EFtransform}, \ref{eq:EFtransform1}) this can be rewritten in terms of ingoing and outgoing EF coordinates:
\begin{align}
    \Phi _{\omega k} & \sim \frac{1}{{\sqrt {r}}} \left\{ 
    A_{\omega k} \exp \left[ -i\left( \omega u  -k{\widehat {\varphi }}_{u} \right)  \right]
    \right. \nonumber 
    \\ & \left.  \qquad 
    + B_{\omega k} \exp \left[ -i\left( \omega v  -k{\widehat {\varphi }}_{v} \right)  \right] \right\}.
\end{align}
The second term corresponds to an ingoing wave with amplitude $|B_{\omega k}|$, and the first term to a wave which is outgoing close to the horizon and has amplitude $|A_{\omega k}|$.
Since we are interested in superradiantly scattered modes, we set $A_{\omega k}=0$ so that the wave is purely ingoing at the horizon.

When the frequency $\omega $ is real, the potential $V_{\omega k}(r)$ (\ref{eq:potential}) is also real and therefore the Wronskian
\begin{equation}
	W_{\omega k} = X_{\omega k}^{*} \frac{dX_{\omega k}}{dr_{*}} - X_{\omega k} \frac{dX_{\omega k}^{*}}{dr_{*}} 
	\label{eq:Wronskian}
\end{equation}
is a constant for any solution $X_{\omega k}$ of the radial equation (\ref{eq:radial}). 
Near the horizon, using (\ref{eq:Xhorizon}) with $A_{\omega k}=0$ we find
\begin{equation}
	W_{\omega k} =-2i{\widetilde {\omega }} \left| B_{\omega k}\right| ^{2}  .
\label{eq:Wronskianhorizon}
\end{equation}
The quantity (\ref{eq:Wronskianhorizon}) is proportional to the radial component of the current \eqref{eq:current} evaluated near the horizon for a scalar field mode having a radial function of the form (\ref{eq:Xhorizon}) with real frequency and $A_{\omega k}=0$, namely 
\begin{equation}
  \left. J^{r}  \right| _{r\rightarrow r_{h}} = -\frac{iq}{r_{h}}W_{\omega k} = -\frac{2q{\widetilde {\omega }}}{r_{h}} \left| B_{\omega k}\right| ^{2}. 
  \label{eq:currenthorizon}
\end{equation}
The fact that the Wronskian is a constant reflects the conservation of the current $J^{\mu }$. 

The value of $W_{\omega k}$ as $r\rightarrow \infty $  depends on the form of the radial function $X_{\omega k}(r)$ in this regime. 
Consider first the solution (\ref{eq:Xinfinity1}) valid when $4\mu ^{2}>-1$. In this case we have
\begin{equation}
	W_{\omega k} = \frac{2i}{\ell ^{2}} \Im (C_{\omega k}^{*} D_{\omega k}) {\sqrt {1+4\mu ^{2}}} ,
	\label{eq:Wronskianinfinity1} 
\end{equation}
where $\Im $ denotes the imaginary part. 
Therefore, if we specify Dirichlet boundary conditions (for which $D_{\omega k}=0$) 
or Neumann boundary conditions (for which $C_{\omega k}=0$), equating (\ref{eq:Wronskianhorizon}, \ref{eq:Wronskianinfinity1}) gives that $\left| B_{\omega k}\right| ^{2}=0$, and it is therefore not possible to have a purely ingoing mode with real frequency at the horizon, generalizing the result of \cite{Ortiz:2011wd} to the charged case.

When $-1<4\mu ^{2}<3$, $\mu ^{2}\neq 0$, Dirichlet and Neumann boundary conditions are not the only possibility, we can also impose Robin boundary conditions for which both $C_{\omega k}$ and $D_{\omega k}$ are nonzero.  
Using the parameterization (\ref{eq:XinfinityRobin}), we have $C_{\omega k} = E_{\omega k } \cos \zeta $ and $D_{\omega k } = E_{\omega k} \sin \zeta  $ and the Wronskian (\ref{eq:Wronskianinfinity1}) becomes
\begin{equation}
	W_{\omega k} = \frac{2i}{\ell ^{2}} \Im \left( \left|E_{\omega k}\right|^{2} \cos \zeta \sin \zeta  \right) {\sqrt {1+4\mu ^{2}}} =0.
\end{equation}
Therefore, when Robin boundary conditions are applied, it is again the case that $\left| B_{\omega k}\right| ^{2}=0$.

There are two special cases which need to be considered separately. First, when $4\mu ^{2}=-1$, the radial function $X_{\omega k}(r)$ has the form (\ref{eq:Xinfinity2}) as $r\rightarrow \infty $, whence 
\begin{equation}
	W_{\omega k} =\frac{2i}{\ell ^{2}}\Im (C_{\omega k}^{*} D_{\omega k}) .
	\label{eq:Wronskianinfinity2}
\end{equation}
Our conclusions are however unchanged: (\ref{eq:Wronskianinfinity2}) vanishes for Dirichlet, Neumann and Robin boundary conditions and \\ $\left| B_{\omega k}\right| ^{2}=0$.

Finally we have the case $4\mu ^{2}=0$, for which the radial function takes the form (\ref{eq:Xinfinity3}) as $r\rightarrow \infty $, with $D_{\omega k}=0$ to ensure square integrability.
In this situation the Wronskian tends to zero as $r\rightarrow \infty $, so that again it must be the case that $\left|  B_{\omega k} \right|^{2}=0$.

The inclusion of a scalar field charge has made no difference to the analysis of 
the Wronskian. 
In particular, we find that 
for charged scalar field modes having real frequency, any mode which is purely ingoing at the event horizon vanishes identically, 
irrespective of the boundary conditions. 
In particular, this means that there can be no superradiant scattering for modes with real frequency.

Similar results were found in \cite{Katagiri:2020mvm} for a charged scalar field on a four-dimensional Reissner-Nordstr\"om-anti-de Sitter black hole. 
As in \cite{Katagiri:2020mvm}, the Wronskian (and hence the radial component of the current $J^{r}$ (\ref{eq:current})) vanishes on the boundary for Robin boundary conditions. 
As a result, the radial component of the current $J^{r}$ (\ref{eq:currenthorizon}) must also vanish on the event horizon.
Therefore there can be no net positive flux from the horizon, and no superradiant scattering.

Our results in this section are in accordance with those of \cite{Dappiaggi:2017pbe} for the neutral scalar field. 
In that case there are superradiantly scattered modes when Robin boundary conditions are applied, but these modes have complex frequencies.  
Our analysis in this subsection is valid only when the modes have real frequencies. In the next subsection we therefore examine the possibility of superradiant scattering for modes with complex frequencies using a different strategy.

\subsection{Fluxes of energy and charge down the horizon}
\label{sec:flux}

If the frequency $\omega $ is complex, it is no longer the case that the Wronskian (\ref{eq:Wronskian}) is a constant and the approach taken in Subsection~\ref{sec:Wronskian} is not applicable. 
In this subsection, we therefore take an alternative perspective to investigate whet-her there are superradiantly scattered modes having complex frequency $\omega $.
Following, for example \cite{Teukolsky:1974yv,Dappiaggi:2017pbe}, we consider the energy flux down the event horizon due to an ingoing mode.
For the remainder of this paper the frequency $\omega $ will be complex.

Consider an ingoing mode given, in terms of ingoing EF coordinates (\ref{eq:EFcoordinates}), by
\begin{align}
	\Phi _{\omega k} 
		= \frac{B_{\omega k}}{{\sqrt {r_{h}}}}
			\exp \left[ -i\left( \omega v  -k{\widehat {\varphi }}_{v} \right)  \right] ,
   \label{eq:ingoing1}
\end{align}
where $B_{\omega k}$ is a complex constant,
and whose radial part is therefore 
\begin{equation}
X_{\omega k}(r) \sim e^{-i{\widetilde {\omega }}r_{*}} {\mbox { as $r_{*}\rightarrow -\infty$.}}
\label{eq:ingoing}
\end{equation}In terms of the Killing vectors $\xi = \partial _{v}$ and $\chi = \partial _{v}+\Omega /\ell \partial _{{\widehat {\varphi}}_{v}}$, where $\Omega /\ell $ is the angular speed of the event horizon, the flux of energy down the black hole is \cite{Dappiaggi:2017pbe}
\begin{align}
	{\mathcal {F}}_{E} = & ~ \int _{0}^{2\pi }d{\widehat {\varphi }}_{v} \, r_{h} \chi _{\mu }T^{\mu }_{\nu }\xi ^{\nu }
	=  2\pi r_{h}\left. T^{r}_{t}\right|_{r=r_{h}} {\sqrt {1-\Omega ^{2}}},
	\label{eq:flux}
\end{align}
where the stress-energy tensor for the charged scalar field is given by (\ref{eq:SET}).
Evaluating the required components of the stress-energy tensor (\ref{eq:SET}) gives
\begin{equation}
	\frac{{\mathcal {F}}_{E}}{F} = \Re (\omega)  \left[
	\Re (\omega )-\frac{k\Omega}{\ell }  \right] + \Im (\omega )^2 ,
 \label{eq:flux1}
\end{equation}
where 
\begin{equation}
    F = 2\pi r_{h}\left|B_{\omega k} \right|^{2} e^{2v\Im (\omega )} .
\end{equation}
Thus the flux of energy down the horizon due to an ingoing mode will be positive unless 
\begin{equation}
	\Re (\omega)  \left[
	\Re (\omega )-\frac{k\Omega}{\ell }  \right] + \Im (\omega )^2< 0.
	\label{eq:SRcriterion}
\end{equation}
This is exactly the same condition for superradiant scattering as in the neutral case \cite{Dappiaggi:2017pbe}.  

Following the analysis in \cite{Katagiri:2020mvm}, we also consider the radial component of the scalar current (\ref{eq:current}). For an ingoing mode of the form (\ref{eq:ingoing1}), near the horizon this is given by
\begin{equation}
 \left. J^{r}  \right| _{r\rightarrow r_{h}}  = -\frac{2q}{r_{h}{\sqrt{1-\Omega ^{2}}}}\left|  B _{\omega k} \right| ^{2} \left[ \Re (\omega )-\frac{k\Omega}{\ell }   \right]  e^{2v\Im (\omega )} . 
 \label{eq:currenthor}
\end{equation}
Assuming that $q>0$, this will be negative if $\Re(\omega)>k\Omega \ell ^{-1}$, corresponding to an ingoing flux of charge.
However, if  \\ $\Re(\omega )<k\Omega \ell ^{-1}$ and $q>0$, we have $J^{r}>0$, and there is an outgoing flux of charge. 
An ingoing mode with $q>0$ and satisfying the criterion (\ref{eq:SRcriterion}) is therefore extracting both charge and energy from the black hole.
In this situation the scalar field mode has been superradiantly scattered. 

For a nonrotating charged black hole we have
\begin{equation}
\frac{{\mathcal {F}}_{E, \Omega =0}}{F}=  \Re (\omega) ^{2} + \Im (\omega )^2  \ge 0 	
\end{equation}
and 
\begin{equation}
 \left. J^{r}  \right| _{r\rightarrow r_{h}} = -\frac{2q}{r_{h}{\sqrt{1-\Omega ^{2}}}}\left|  B _{\omega k} \right| ^{2} \Re (\omega ) e^{2v\Im (\omega )}  <0  
\end{equation}
if $q>0$ and $\Re(\omega )>0$.
Therefore an ingoing mode has ingoing fluxes of both energy and charge and there is no superradiant scattering for a charged scalar field on a static, charged BTZ black hole, irrespective of the boundary conditions applied to the field.

This result is in contrast to that for a charged scalar field on a four-dimensional, nonrotating, asymptotically flat Reis-sner-Nordstr\"om black hole \cite{Katagiri:2020mvm}, for which superradiant scattering can occur. The key difference between these two situations is that, for charged scalar field scattering on a Reissner-Nordstr\"om black hole, the electrostatic potential at the horizon  plays the same role as $k\Omega \ell ^{-1}$ does in (\ref{eq:currenthor}), resulting in a range of ingoing mode frequencies for which superradiant scattering is possible.
In our case, the electromagnetic potential (\ref{eq:rotatinggaugepot}) vanishes on the horizon.
Therefore the black hole charge $Q$ does not appear in the fluxes of energy (\ref{eq:flux1}) or charge (\ref{eq:currenthor}) at the event horizon.

\section{Superradiant scattering on charged rotating BTZ black holes}
\label{sec:rotating} 

We now provide evidence for the existence of superradiantly scattered modes satisfying the condition (\ref{eq:SRcriterion}), using a numerical method. We restrict attention to the regime $-1<4\mu ^{2}<3$, $\mu ^{2}\neq 0$ for which there is a choice of boundary conditions that can be applied to the scalar field at infinity. 

Our numerical method involves integrating the radial eq-uation (\ref{eq:radial}) to find complex frequencies $\omega $ for which the radial function $X_{\omega \ell }(r)$ satisfies ingoing boundary conditions (\ref{eq:ingoing}) at the horizon and Robin boundary conditions (\ref{eq:XinfinityRobin}) at infinity. 
In other words, we are seeking quasi-normal modes (QNMs). 
There are many methods in the literature for the accurate computation of QNM frequencies (see, for example \cite{Kokkotas:1999bd,Berti:2009kk,Konoplya:2011qq,Cho:2011sf} for reviews and \cite{Birmingham:2001hc,Cardoso:2001hn,Crisostomo:2004hj,Decanini:2009dn,Zhang:2011ec,Kandemir:2016pjf,Panotopoulos:2018can,Rincon:2018sgd,Dias:2019ery,Singha:2022bvr,Katagiri:2022qje,DalBoscoFontana:2023syy,Fontana:2023dix} for a selection of references concerning QNMs on BTZ black holes). 
However, the form of the potential (\ref{eq:potential}) (in particular, the presence of the nonanalytic $\ln \left( \frac{r}{r_{h}}\right) $ term)  hinders implementing these in our situation.  
Our aims in this section are rather less ambitious than the high-precision computation of QNM frequencies. 
Instead, we are looking for numerical evidence for the existence of superradiantly scattered modes, and some qualitative information about the energy flux (\ref{eq:flux1}) for these modes.
With this in mind, we employ a rather naive direct integration method, which is sufficiently accurate for our purposes for modes lying in the region (\ref{eq:SRcriterion}) for which superradiant scattering is possible. Our computations are described in Appendix \ref{sec:method}  and are implemented in {\tt {MATHEMATICA}}.

According to (\ref{eq:SRcriterion}), superradiantly scattered modes can exist only for frequencies having a real part $\Re (\omega )$ satisfying $0<\Re (\omega )<k\Omega /\ell $ for $k>0$.
We therefore consider only frequencies whose real parts lie in this interval. 
For a general frequency $\omega $, the constants $C_{\omega k}$ and $D_{\omega k}$ (\ref{eq:Xinfinity1}) 
describing the behaviour of the scalar field mode at infinity will be complex.
To apply Robin boundary conditions (\ref{eq:XinfinityRobin}), we require the ratio $D_{\omega k}/C_{\omega k}$ to be real, which will only occur at particular frequencies.
When the imaginary part of $D_{\omega k}/C_{\omega k}$ vanishes,
we determine the parameter $\zeta $ (\ref{eq:XinfinityRobin}) governing the Robin boundary conditions from
\begin{equation}
\zeta = \arctan \left( \frac{D_{\omega k}}{C_{\omega k}} \right) ,
\label{eq:Zeta}
\end{equation}
and the energy flux ${\mathcal {F}}_{E}/F$ using (\ref{eq:flux1}).
We mostly take a branch of the $\arctan $ function such that $\zeta \in [0, \pi )$; however in part of our analysis it will be helpful to consider instead $\zeta \in (-\frac{\pi }{2},\frac{\pi }{2}]$.

We find that our numerical method yields satisfactory results only when either the scalar field charge vanishes ($q=0$) or for reasonably large values of at least one of the charges $|Q|$, $q$. 
In order to obtain good results for a wider range of values of the charges, and for 
nonsuperradiantly scattered modes, a more sophisticated method would be needed to find the QNM frequencies.
However, our method is sufficiently accurate to give a selection of superradiantly scattered modes which enables us to qualitatively explore the effect of black hole and/or scalar field charge on superradiant scattering.
Below we consider first the energy flux (\ref{eq:flux1}) from superradiantly scattered modes and then the real and imaginary parts of the mode frequencies.

For our discussion of superradiantly scattered modes, it will be important to recall, from (\ref{eq:charge}), that the black hole charge ${\widetilde {Q}}$ has the opposite sign to the parameter $Q$.
Therefore, if $qQ>0$, we have $q{\widetilde {Q}}<0$ and the scalar field charge has the opposite sign to the black hole charge, while if $qQ<0$, then $q{\widetilde {Q}}>0$ and the scalar field charge has the same sign as the black hole charge. 
The radial equation (\ref{eq:radial}) depends only on the product $qQ$ (explicit in the first term in the potential $V_{\omega k}$ (\ref{eq:potential})) and $Q^{2}$ (which appears in the metric function $N(r)$ (\ref{eq:chargedlapse}) and its derivative).
Changing the sign of the black hole charge parameter $Q$ (with the scalar field charge $q$ fixed) will only impact the second term in the square brackets in the first line of the potential (\ref{eq:potential}).
All other terms in the potential are unaffected on changing the sign of $Q$. 
In this section we therefore consider positive scalar field charge $q>0$ and both positive and negative values of the parameter $Q$.

\subsection{Energy flux}
\label{sec:energy}

\begin{figure}[th!]
        \includegraphics[width=\linewidth]{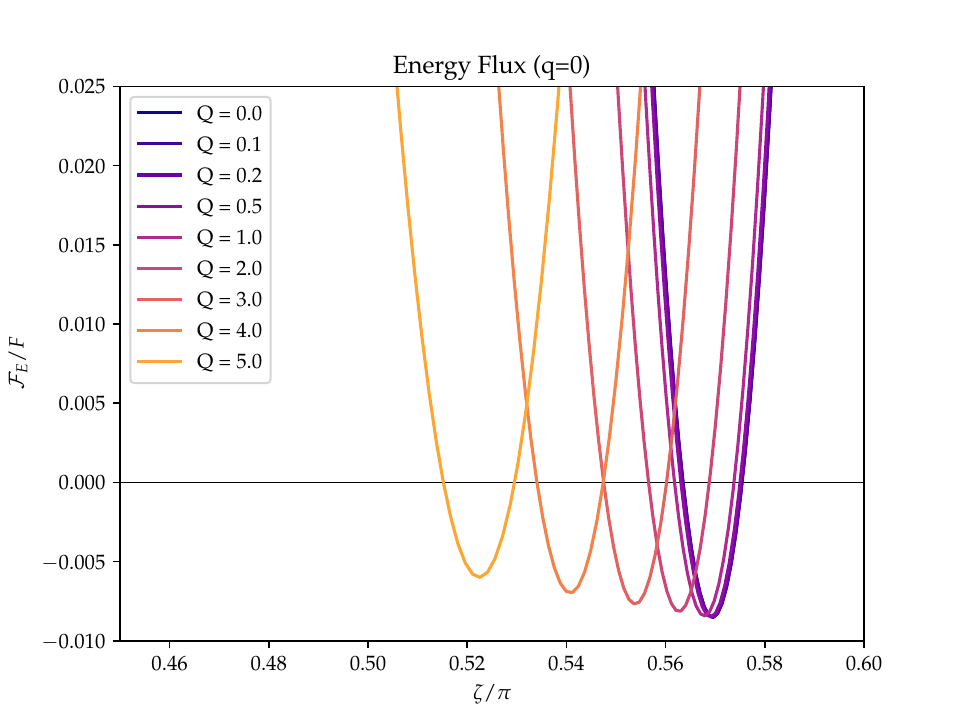}%
\caption{Energy flux ${\mathcal {F}}_{E}/F$ (\ref{eq:flux1}) for superradiantly scattered neutral scalar field modes as a function of the Robin parameter $\zeta $ (\ref{eq:Zeta}).  A negative energy flux corresponds to superradiant scattering. 
We have fixed $m^{2}=-0.65$, $\ell =1$, $M=16$, $\Omega = 0.6$, $q=0$ and $k=1$. 
The values of the black hole charge parameter $Q$ are as given in the legends.}
\label{fig:Resultsq0}
\end{figure}

\begin{figure*}[th!]
    \subfloat[]{%
        \includegraphics[width=.50\linewidth]{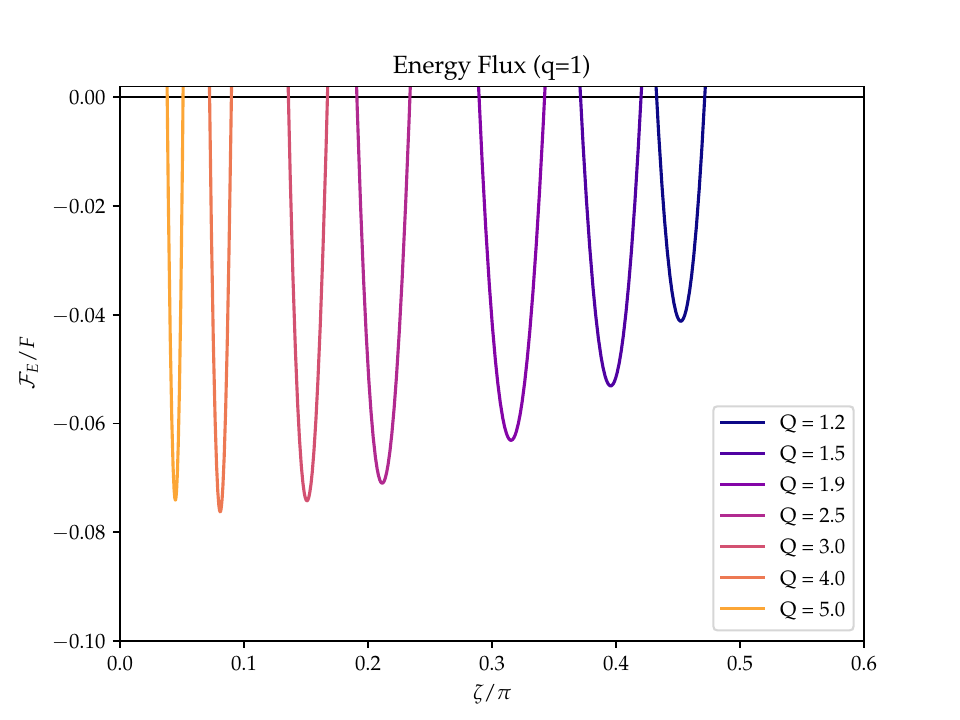}%
        \label{subfig:a}%
    } \hfill
    \subfloat[]{%
        \includegraphics[width=.50\linewidth]{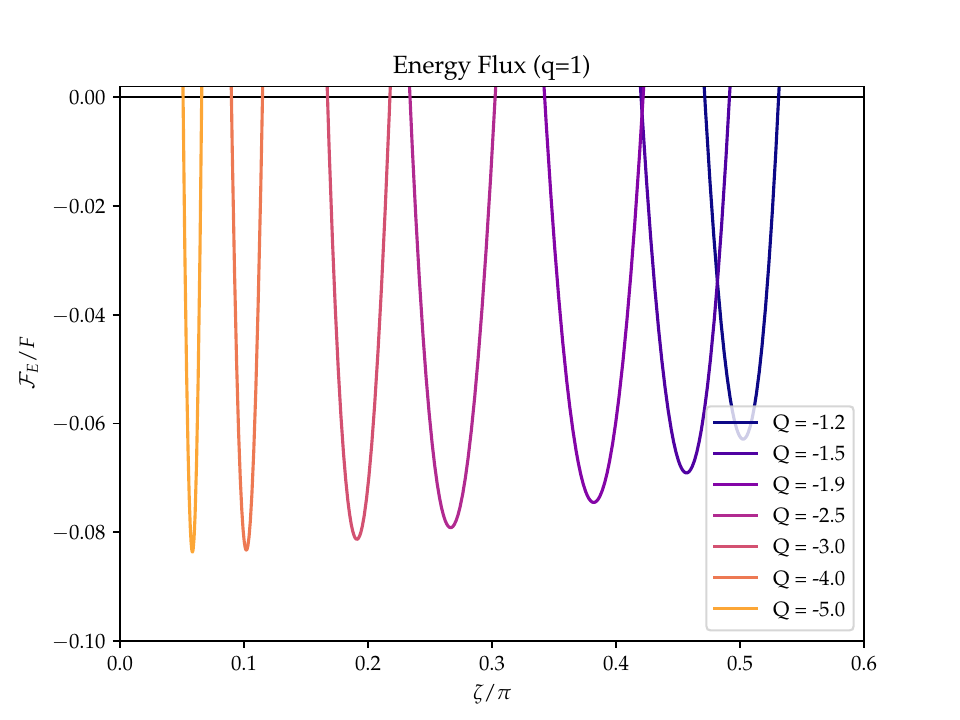}%
        \label{subfig:b}%
        }
    \\
    \subfloat[]{%
        \includegraphics[width=.50\linewidth]{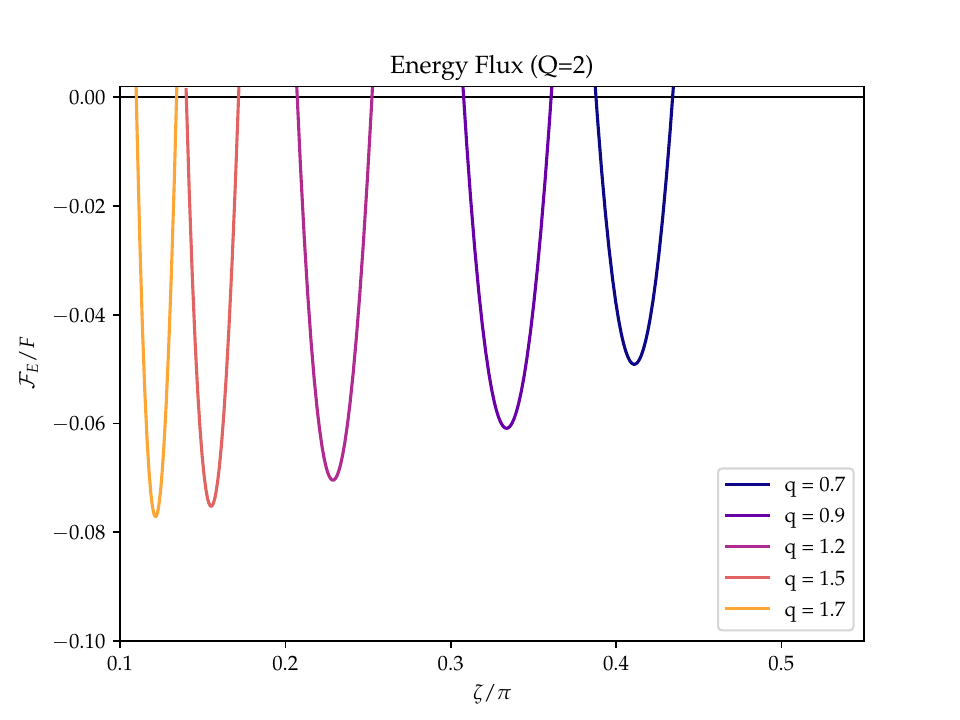}%
        \label{subfig:c}%
    } \hfill 
     \subfloat[]{%
        \includegraphics[width=.50\linewidth]{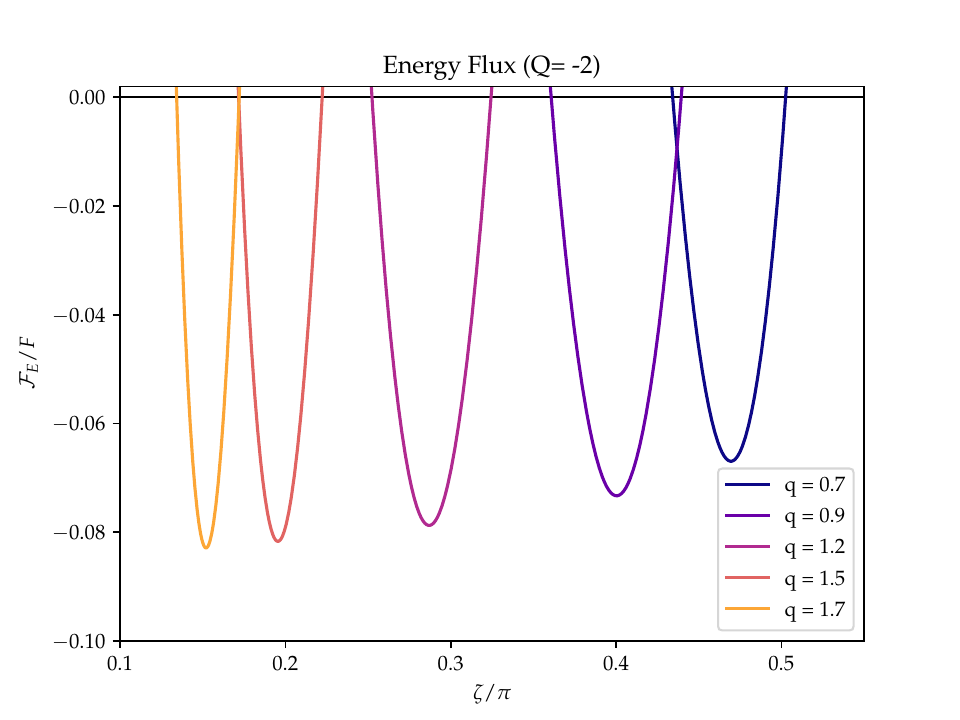}%
        \label{subfig:d}%
    }
    \\
\caption{Energy flux ${\mathcal {F}}_{E}/F$ (\ref{eq:flux1}) for superradiantly scattered charged scalar field modes as a function of the Robin parameter $\zeta $ (\ref{eq:Zeta}).  A negative energy flux corresponds to superradiant scattering. 
We have fixed $m^{2}=-0.65$, $\ell =1$, $M=16$, $\Omega =0.6$ and $k=1$. 
The values of the black hole charge parameter $Q$ and scalar field charge $q$ are as given in the legends.} 
\label{fig:Results}
\end{figure*}

\begin{figure*}[th!]
    \subfloat[]{%
        \includegraphics[width=.5\linewidth]{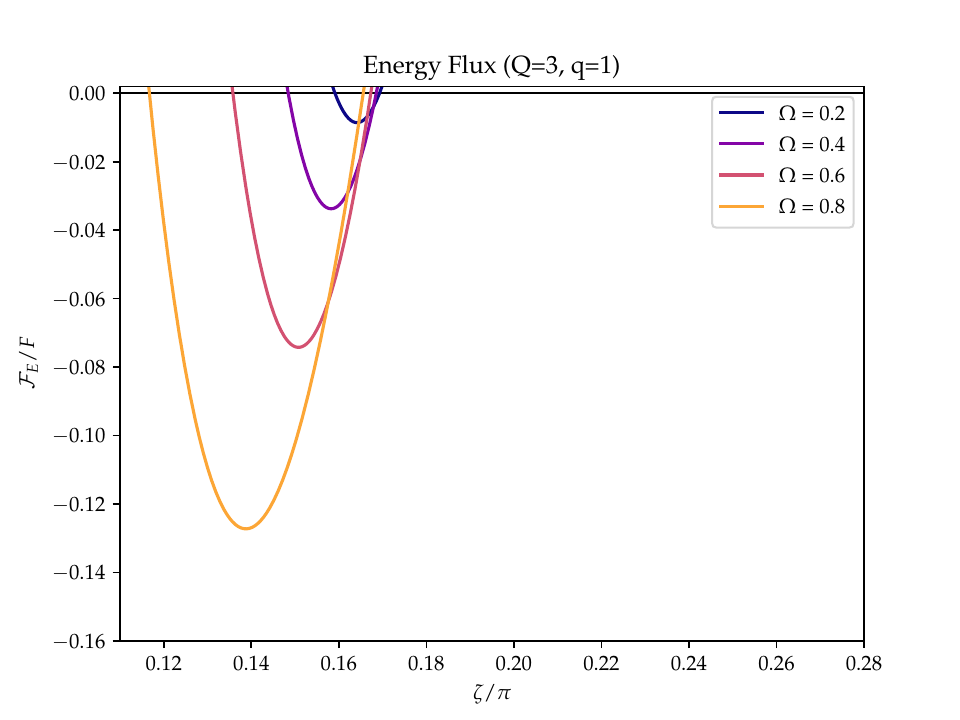}%
        \label{subfig:Rotation1}%
    }\hfill
    \subfloat[]{%
        \includegraphics[width=.5\linewidth]{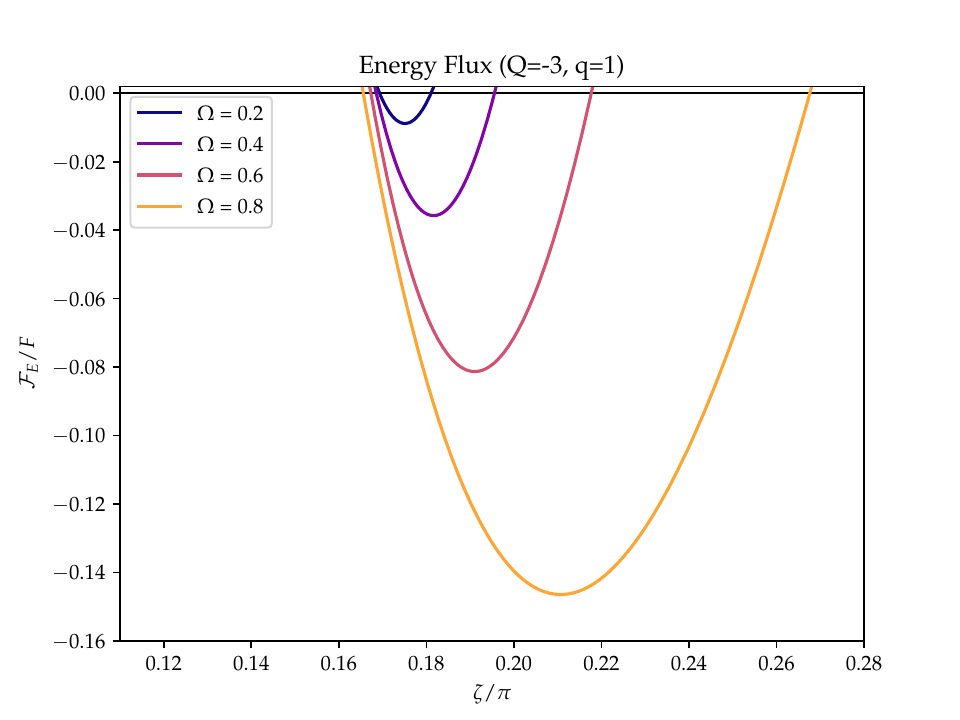}%
        \label{subfig:Rotation2}%
    } 
\caption{Energy flux ${\mathcal {F}}_{E}/F$ (\ref{eq:flux1}) for superradiantly scattered charged scalar field modes as a function of the Robin parameter $\zeta $ (\ref{eq:Zeta}).  A negative energy flux corresponds to superradiant scattering. 
We have set $m^{2}=-0.65$, $\ell =1$, $M=16$ and $k=1$.  The black hole charge parameter is fixed to be $Q=\pm 3$, and the scalar field charge is $q=1$. A selection of values of the rotation parameter $\Omega $ are considered.}
\label{fig:Rotation}
\end{figure*}

\begin{figure*}[th!]
    \subfloat[]{%
        \includegraphics[width=.5\linewidth]{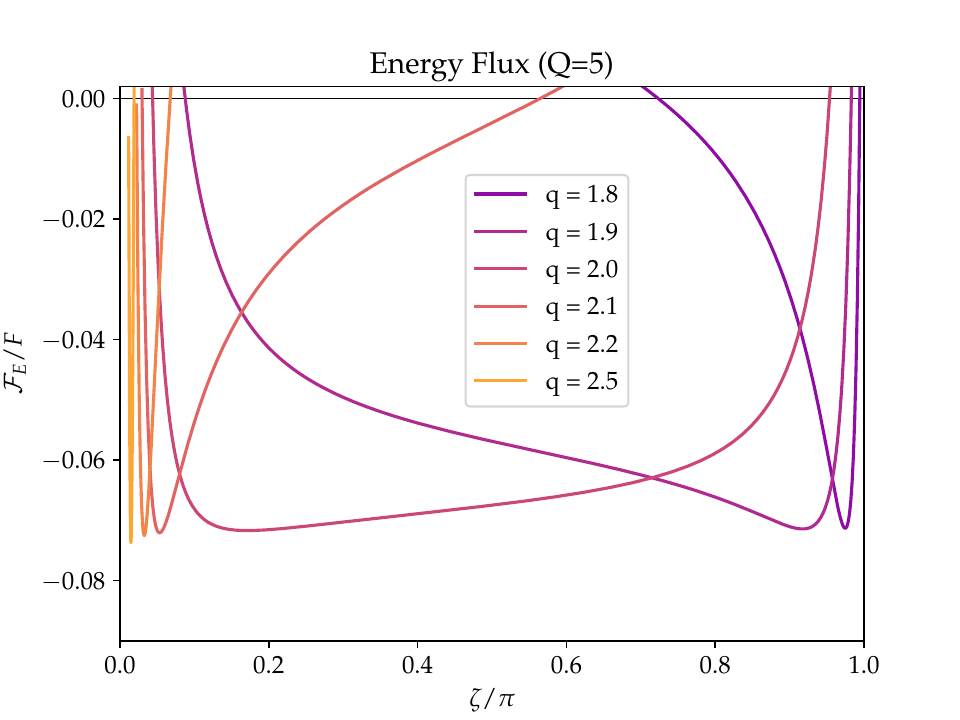}%
        \label{subfig:FirstBranch}%
    }\hfill
    \subfloat[]{%
        \includegraphics[width=.5\linewidth]{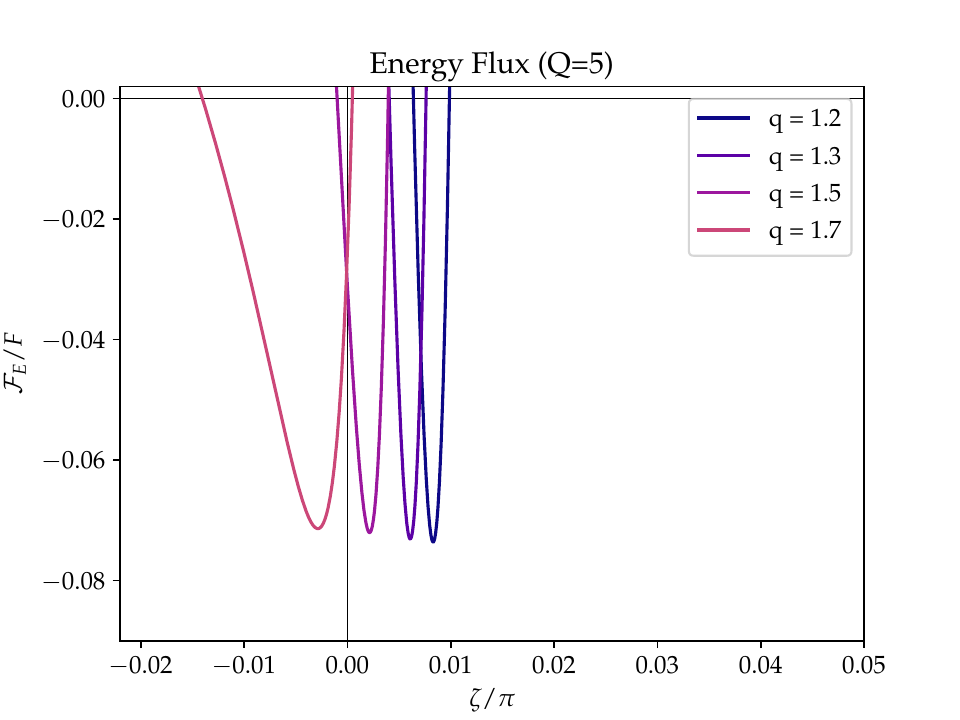}%
        \label{subfig:SecondBranch}%
    } 
    \\
    \subfloat[]{%
        \includegraphics[width=.5\linewidth]{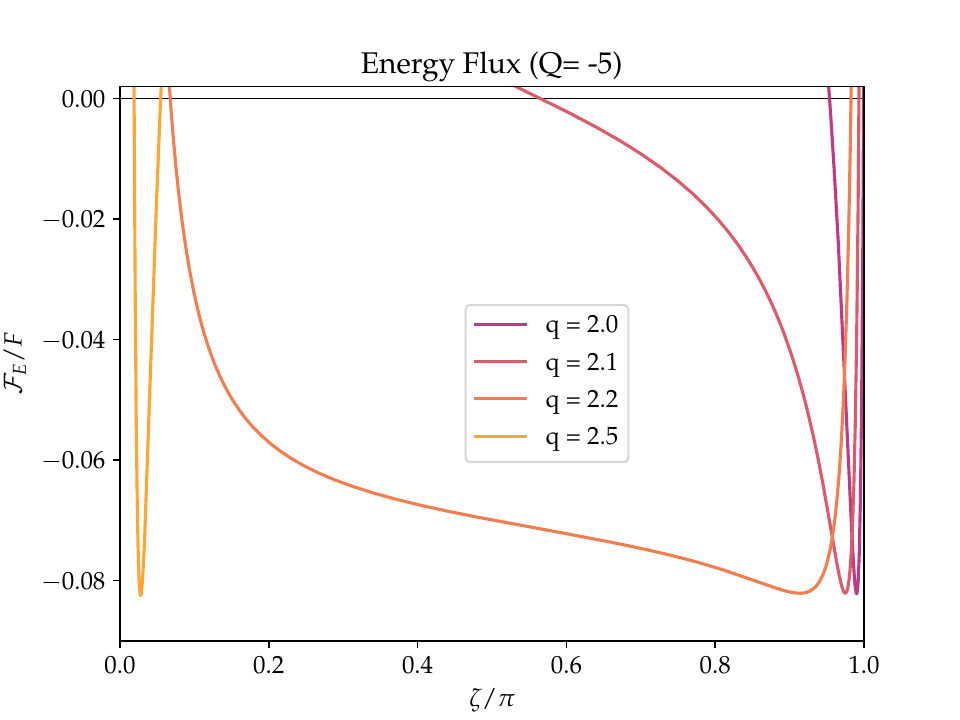}%
        \label{subfig:FirstBranch1}%
    }\hfill
    \subfloat[]{%
        \includegraphics[width=.5\linewidth]{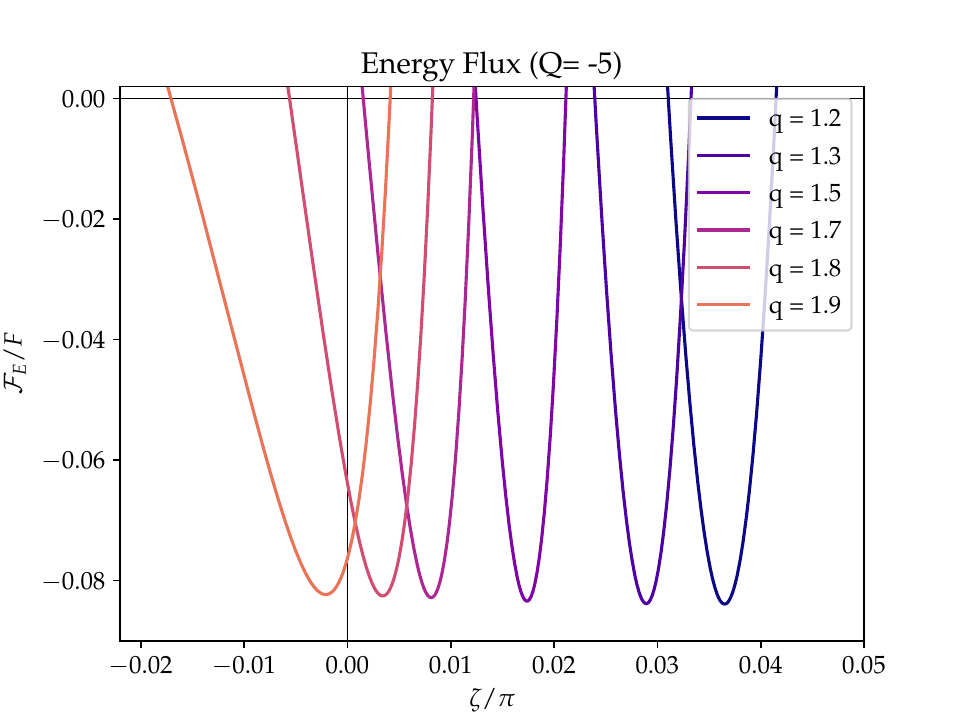}%
        \label{subfig:SecondBranch1}%
    } 
\caption{Energy flux ${\mathcal {F}}_{E}/F$ (\ref{eq:flux1}) for superradiantly scattered charged scalar field modes as a function of the Robin parameter $\zeta $ (\ref{eq:Zeta}).  A negative energy flux corresponds to superradiant scattering. 
We have set $m^{2}=-0.65$, $\ell =1$, $M=16$, $\Omega = 0.6$ and $k=1$.  The black hole charge parameter is fixed to be $Q=\pm 5$, and a selection of values of the scalar field charge $q$ are considered.}
\label{fig:Branches}
\end{figure*}

In Figures \ref{fig:Resultsq0}--\ref{fig:Branches} we present our numerical results providing evidence for the existence of superradiant scattering for a charged scalar field on a charged, rotating BTZ black hole.
To aid comparison with the results for a neutral scalar field in \cite{Dappiaggi:2017pbe}, we set $m^{2}=-0.65$, $\ell =1$, $M=16$ (which corresponds to Figure 2 in \cite{Dappiaggi:2017pbe}) and consider only modes with azimuthal quantum number $k=1$.
In all our plots we show the energy flux ${\mathcal {F}}_{E}/F$ (\ref{eq:flux1}) as a function of the Robin parameter $\zeta $ (\ref{eq:Zeta}). 
A negative energy flux corresponds to superradiant scattering.

We begin by setting the scalar field charge $q=0$, see Figure \ref{fig:Resultsq0}, where we have also fixed the rotation parameter $\Omega = 0.6$ and varied the black hole charge parameter $Q$. 
For a neutral scalar field with $q=0$, the radial equation (\ref{eq:radial}) does not depend on the sign of the charge parameter $Q$ so it is sufficient to consider $Q\ge 0$.
When $Q=0$, we reproduce the results in \cite{Dappiaggi:2017pbe}, which provides verification of our numerical method. 
QNM for $Q\neq 0 $ and $q=0$ were studied in \cite{Tang:2016vmu}, although the focus in that work was the mode frequencies rather than the energy flux, as is the case here.
It can be seen in Figure \ref{fig:Resultsq0} that, as the black hole charge parameter $Q$ increases, the maximum magnitude of the energy flux ${\mathcal {F}}_{E}/F$  in superradiantly scattered modes decreases, and that superradiantly scattered modes exist for smaller values of the Robin parameter $\zeta $. 
The effects of superradiant scattering are small in this situation: there is only a narrow interval of values of $\zeta $ for which there are superradiantly scattered modes, and the resulting fluxes of energy have small magnitudes. 
We also note that all values of the Robin parameter $\zeta $ for the superradiantly scattered modes when $q=0$ shown in Figure \ref{fig:Resultsq0} are greater than $\pi /2$, the value corresponding to Neumann boundary conditions.

We now examine the effect of the scalar field charge on the energy flux for superradiantly scattered modes.

In Figure \ref{fig:Results}(a--b) we set the scalar field charge $q=1$ and consider a selection of values of the black hole charge parameter $Q$, again for fixed rotation parameter $\Omega = 0.6$.
We have $qQ>0$ in Figure~\ref{fig:Results}(a) and $qQ<0$ in Figure~\ref{fig:Results}(b).
For fixed $Q$, superradiantly scattered modes exist only in a narrow interval of values of the Robin parameter $\zeta $, with the width of this interval decreasing as $|Q|$ increases.
The widths of these intervals are slightly greater when $qQ<0$ compared to $qQ>0$.
As $Q$ varies, the possible values of $\zeta $ for which there are superradiantly scattered modes is much broader than in the case $q=0$, and we find superradiantly scattered modes with boundary conditions close to Dirichlet ($\zeta = 0$) when $|Q|$ is large.
The magnitude of the energy flux ${\mathcal {F}}_{E}/F$ for the superradiantly scattered modes with $q=1$ in Figure \ref{fig:Results}(a--b) is roughly an order of magnitude greater than those in Figure \ref{fig:Resultsq0} for $q=0$, indicating a significant enhancement in superradiant scattering due to the scalar field charge.
The maximum magnitude of the energy flux is slightly larger for superradiantly scattered modes with $qQ<0$ than for the same values of $q$ and $|Q|$ but with $qQ>0$. 
Therefore there is a slight additional enhancement in superradiant scattering when the scalar field charge $q$ has the same sign as the black hole charge ${\widetilde {Q}}$ (\ref{eq:charge}) compared to the case where $q$ has the opposite sign to ${\widetilde {Q}}$.
The values of the Robin parameter $\zeta $ for the superradiantly scattered modes in Figure \ref{fig:Results}(a--b) mostly lie between the Dirichlet value $\zeta =0$ and that for Neumann boundary conditions $\zeta = \frac{\pi }{2}$.

In Figure \ref{fig:Results}(c--d), with the rotation parameter again set to be $\Omega = 0.6$, we fix the black hole charge parameter $Q=\pm 2$ and vary the scalar field charge $q$.
In Figure \ref{fig:Results}(c--d), the interval of values of the Robin parameter $\zeta $ for which superradiantly scattered modes exists shrinks as the scalar field charge $q$ increases, and moves to smaller values of $\zeta $. The widths of these intervals are slightly greater for $Q=-2$ than for $Q=2$ with the same value of $q$. 
At the same time, the maximum magnitude of the energy flux for superradiantly scattered modes  ${\mathcal {F}}_{E}/F$ in Figure \ref{fig:Results}(c--d) increases as $q$ increases for both positive and negative $Q$. 
The maximum magnitude of the energy flux is slightly larger for $Q=-2$ than it is for $Q=2$ with the same value of $q$.
Combining the results in Figures \ref{fig:Results}, we provide evidence that increasing the magnitudes of either the black hole or scalar field charges gives a narrower interval of values of $\zeta $ yielding superradiantly scattered modes, with that interval being closer to Dirichlet boundary conditions.
The maximum magnitude of the energy flux for superradiantly scattered modes shown in Figure \ref{fig:Results} generally increases as either $q$ or $|Q|$ increases.

So far we have studied superradiantly scattered modes with the rotation parameter $\Omega $ fixed. 
In Figure \ref{fig:Rotation} we fix the black hole charge parameter $Q=\pm 3$ and scalar field charge $q=1$, and consider a selection of values of $\Omega $. 
In Figure \ref{fig:Rotation}, increasing the rotation parameter results in large increases in both the width of the interval of values of $\zeta $ for which there are superradiantly scattered modes, and the maximum magnitude of the energy flux for superradiantly scattered modes. 
These effects in Figure \ref{fig:Rotation} are significantly larger than those in Figure \ref{fig:Results} resulting from changing either the scalar field or black hole charges.
We therefore provide evidence that the most important factor influencing superradiant scattering is the rotation of the black hole.

Comparing the results for $Q=\pm 3$ in Figure~\ref{fig:Rotation}, there is again a slight enhancement in the maximum magnitude of the energy flux when $Q<0$ compared to $Q>0$.
The difference in the values of the Robin parameter $\zeta $ for superradiantly scattered modes when $Q=3$ compared to $Q=-3$ is also striking.

We close our discussion of the energy flux by exploring, in Figure \ref{fig:Branches}, some results for a large value of the black hole charge parameter, namely $|Q|=5$, again with the rotation parameter $\Omega = 0.6$. 
In Figure \ref{fig:Branches} we find behaviour which is qualitatively different from that shown in Figure \ref{fig:Results}.

Consider first $Q=5$ and the superradiantly scattered modes shown in Figure \ref{fig:Branches}(a).
For larger values of the scalar field charge $q\ge 2.2$,  in Figure \ref{fig:Branches}(a) we find a narrow interval of values of the Robin parameter $\zeta $ which yield superradiantly scattered modes, and furthermore these values of $\zeta $ lie close to the Dirichlet value $\zeta = 0$, similarly to the results in Figure \ref{fig:Results}(c) for $Q=2$.
However, as $q$ decreases (again in Figure \ref{fig:Branches}(a)), the interval of values of $\zeta $ for which there are superradiantly scattered modes widens significantly, and comprises the majority of the interval $0<\zeta < \pi $. 
In particular, for $q=2.0$, $2.1$ and $2.2$ we show, in Figure \ref{fig:Branches}(a), superradiantly scattered modes for which $\zeta = \frac{\pi }{2}$, corresponding to Neumann boundary conditions.
For $1.8<q<2.2$, Figure \ref{fig:Branches}(a) shows that the value of the Robin parameter $\zeta $ at which the energy flux has its maximum magnitude shifts from a location close to $\zeta = 0$ to a location close to $\zeta = \pi $. 
On decreasing $q$ further, for fixed $q$ we find two ``branches'' of superradiantly scattered modes, one in a neighbourhood of $\zeta = \pi $ and one in a neighbourhood of $\zeta = 0$.
These superradiantly scattered modes are depicted in Figure \ref{fig:Branches}(b), where we have chosen a branch of the $\arctan $ function in (\ref{eq:Zeta}) for which $-\frac{\pi }{2}<\zeta <\frac{\pi }{2}$ instead of $0<\zeta <\pi $ as used elsewhere.
In Figure \ref{fig:Branches}(b), we can see that for $q=1.5$ and $1.7$, there are superradiantly scattered modes for which $\zeta = 0$, corresponding to Dirichlet boundary conditions. 

Superradiantly scattered charged scalar field modes for $Q=-5$ are shown in Figure~\ref{fig:Branches}(c--d). 
Again we find that for large values of the scalar field charge $q$, superradiantly scattered modes lie in a narrow interval of values of the Robin parameter $\zeta$, and these values are close to the Dirichlet value $\zeta =0$.
Decreasing $q$ again results in superradiantly scattered modes in a much larger interval of values of $\zeta $, but the width of the interval of values of $\zeta $ shrinks again at a larger value of $q$ than for the $Q=5$ case.  
In Figure~\ref{fig:Branches}(c), we find superradiantly scattered modes for which $\zeta = \frac{\pi }{2}$  when $q=2.2$.
As $q$ decreases further, as for $Q=5$ we also find two ``branches'' of superradiantly scattered modes, which are shown in Figure~\ref{fig:Branches}(d), where again we use a branch of the $\arctan $ function in (\ref{eq:Zeta}) for which $-\frac{\pi }{2}< \zeta < \frac{\pi }{2}$.
The value of $q$ below which we have two ``branches'' is higher for $Q=-5$ than it is for $Q=5$. 
From Figure~\ref{fig:Branches}(d), we find superradiantly scattered modes for which $\zeta =0$ (corresponding to Dirichlet boundary conditions) when $q=1.8$ and $1.9$.
In Figure~\ref{fig:Branches}, it can also be seen that the maximum value of the energy flux is slightly larger for superradiantly scattered modes with $Q=-5$ than for modes with the same value of the scalar charge $q$ but with $Q=5$.
Comparing Figure~\ref{fig:Branches}(b) and (d), it can be seen that the range  of values of the Robin parameter $\zeta $ is notably different for $Q=-5$ compared to $Q=5$.

We therefore provide evidence that, unlike the situation for a neutral scalar field, for a charged scalar field on a char-ged BTZ black hole background, at least for a small subset of the $(Q,q)$-parameter space, there are superradiantly scattered modes with complex frequencies satisfying either Dirichlet or Neumann boundary conditions at infinity.

\subsection{Mode frequencies}
\label{sec:frequencies}

\begin{figure*}[ht!]
    \subfloat[]{%
        \includegraphics[width=.50\linewidth]{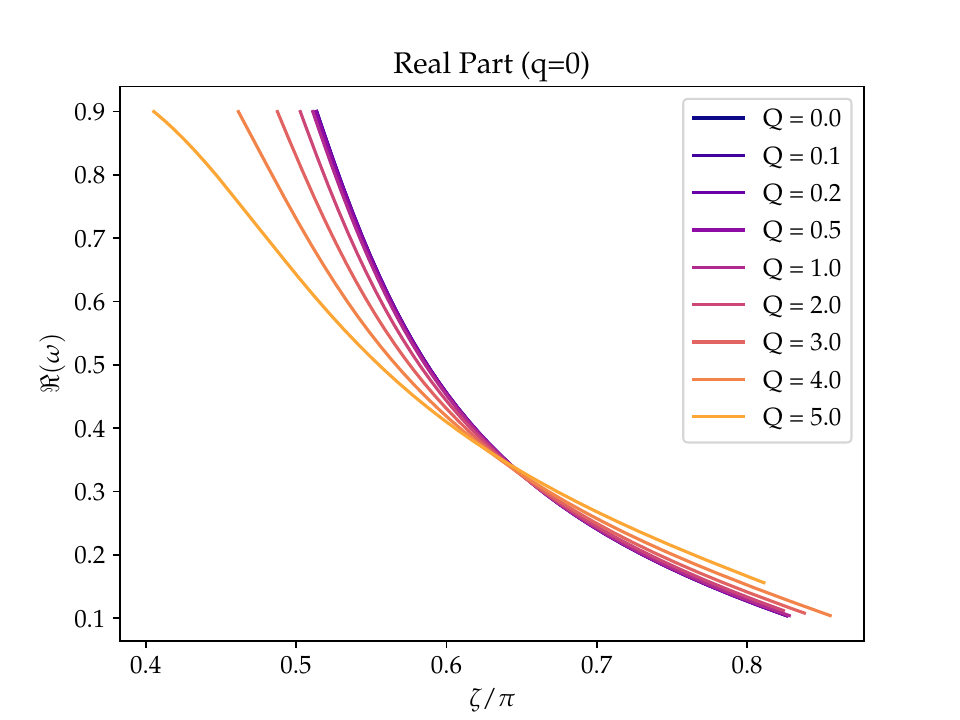}%
        \label{subfig:a2}%
    }\hfill
    \subfloat[]{%
        \includegraphics[width=.50\linewidth]{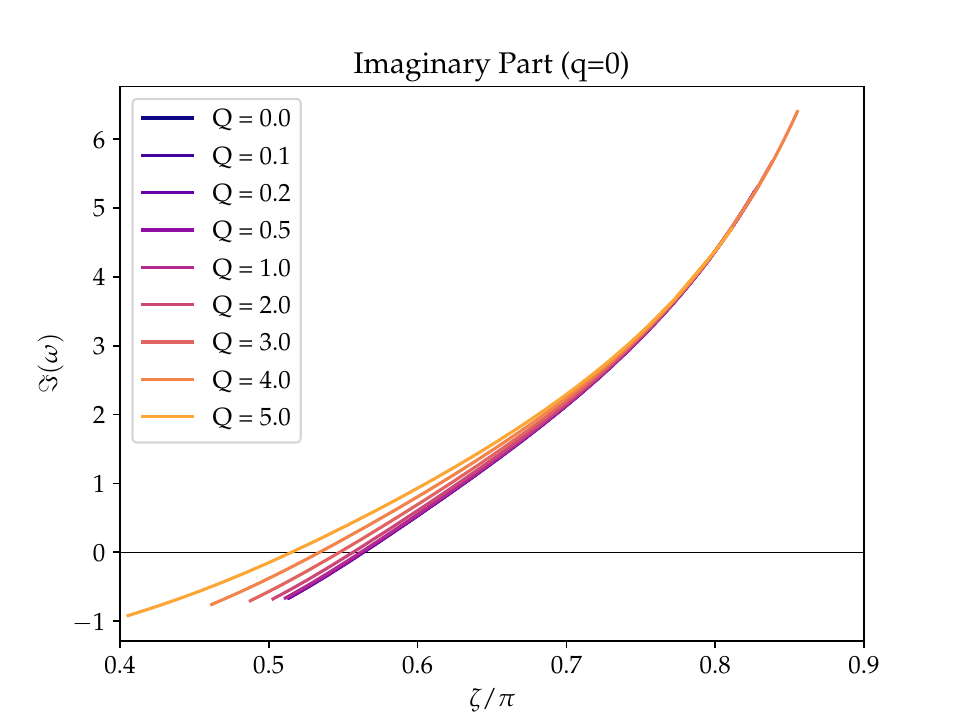}%
        \label{subfig:b2}%
    } 
\caption{Real (left) and imaginary (right) parts of the frequencies of neutral scalar field modes as a function of the Robin parameter $\zeta $ (\ref{eq:Zeta}). We have fixed $m^{2}=-0.65$, $\ell =1$, $M=16$, $\Omega = 0.6$, $q=0$ and $k=1$. 
The values of the black hole charge parameter $Q$ are as given in the legends.}
\label{fig:frequencyq0}
\end{figure*}

\begin{figure*}[ht!]
    \subfloat[]{%
        \includegraphics[width=.50\linewidth]{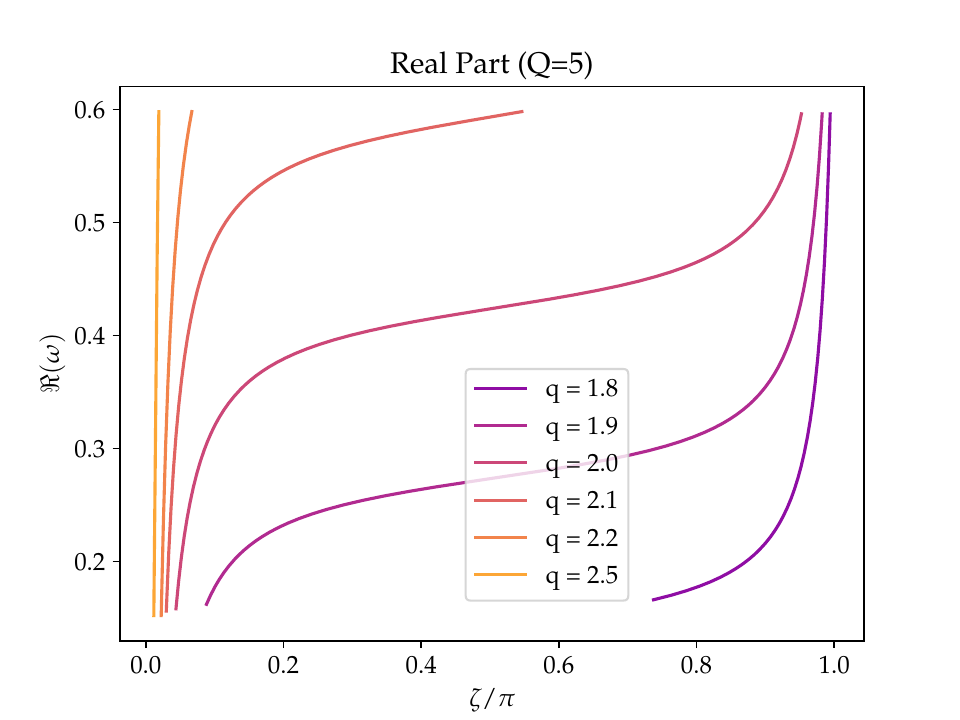}%
        \label{subfig:a3}%
    }\hfill
    \subfloat[]{%
        \includegraphics[width=.50\linewidth]{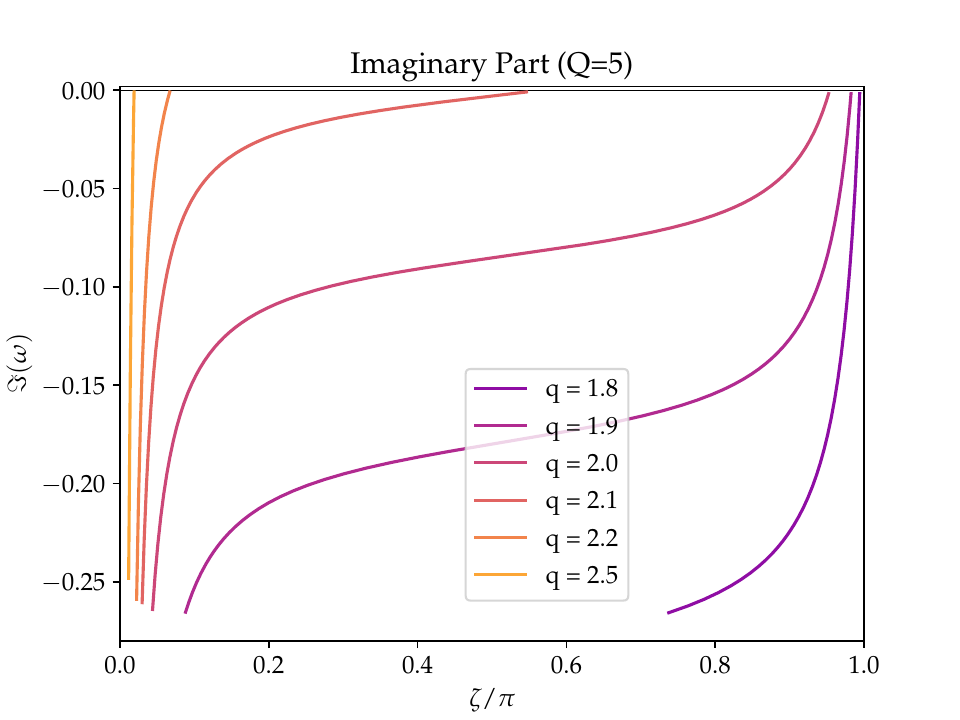}%
        \label{subfig:b3}%
    } 
\\
\subfloat[]{%
        \includegraphics[width=.50\linewidth]{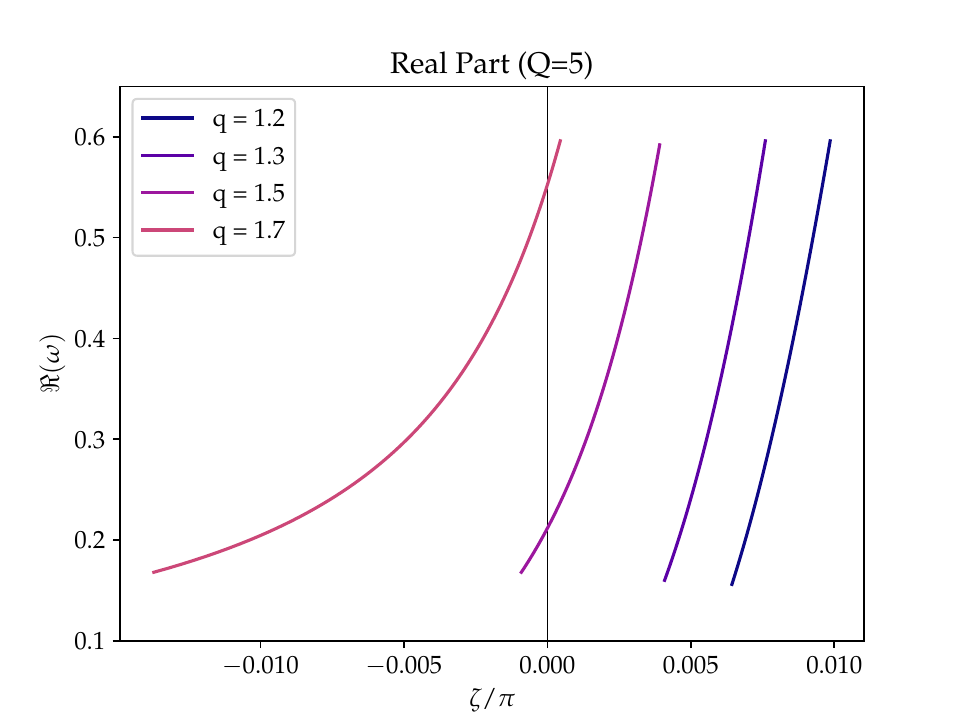}%
        \label{subfig:c3}%
    }\hfill
    \subfloat[]{%
        \includegraphics[width=.50\linewidth]{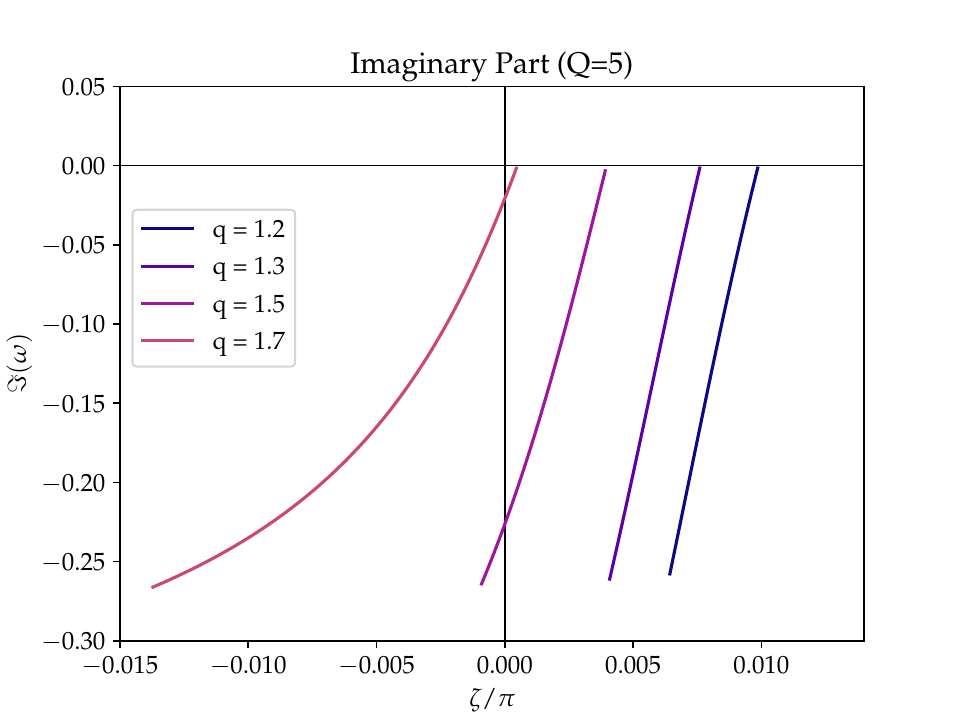}%
        \label{subfig:d3}%
    } 
\caption{Real (left) and imaginary (right) parts of the frequencies of superradiantly scattered charged scalar field modes as a function of the Robin parameter $\zeta $ (\ref{eq:Zeta}). 
We have fixed $m^{2}=-0.65$, $\ell =1$, $M=16$, $\Omega = 0.6$ and $k=1$. 
The black hole charge parameter is fixed to be $Q=5$, and a selection of values of the scalar field charge $q$ are considered.} 
\label{fig:frequencyQ5}
\end{figure*}

\begin{figure*}[ht!]
    \subfloat[]{%
        \includegraphics[width=.50\linewidth]{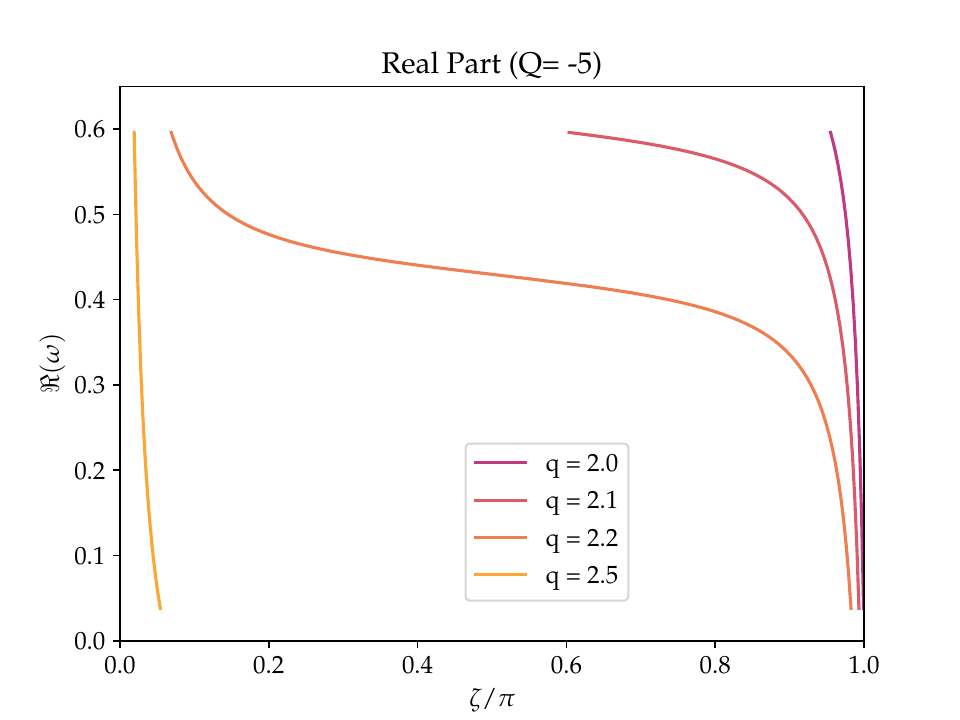}%
        \label{subfig:a31}%
    }\hfill
    \subfloat[]{%
        \includegraphics[width=.50\linewidth]{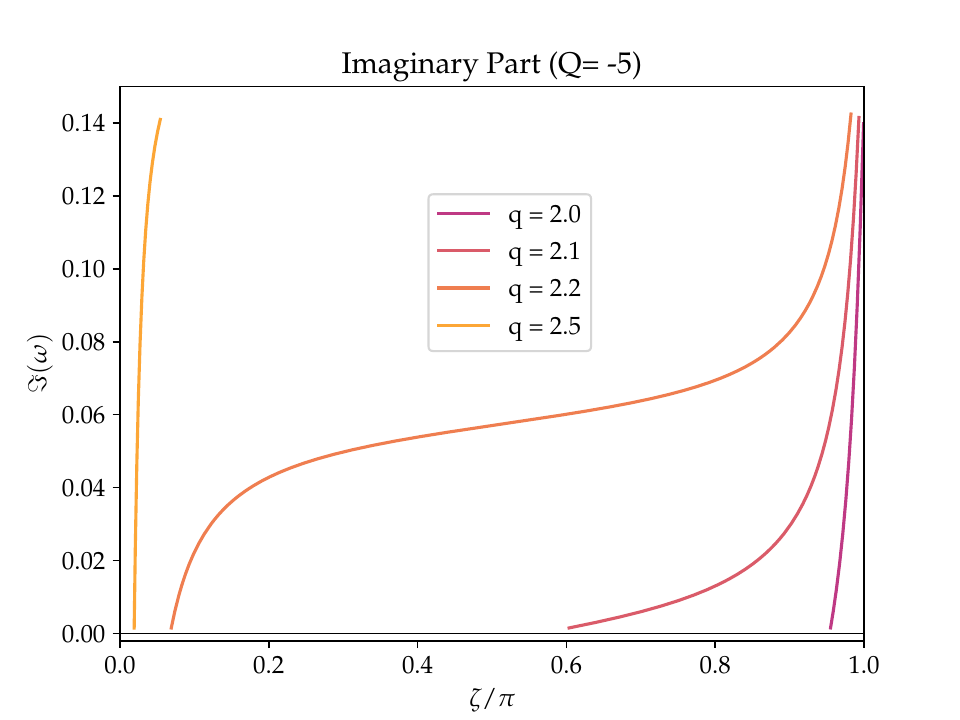}%
        \label{subfig:b31}%
    } 
\\
\subfloat[]{%
        \includegraphics[width=.50\linewidth]{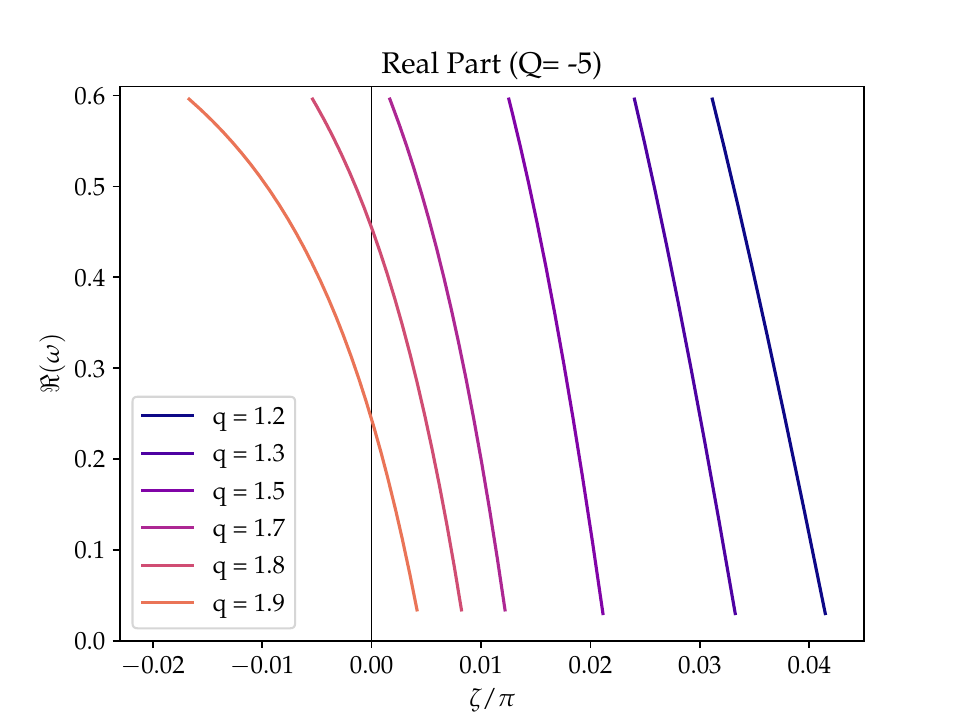}%
        \label{subfig:c31}%
    }\hfill
    \subfloat[]{%
        \includegraphics[width=.50\linewidth]{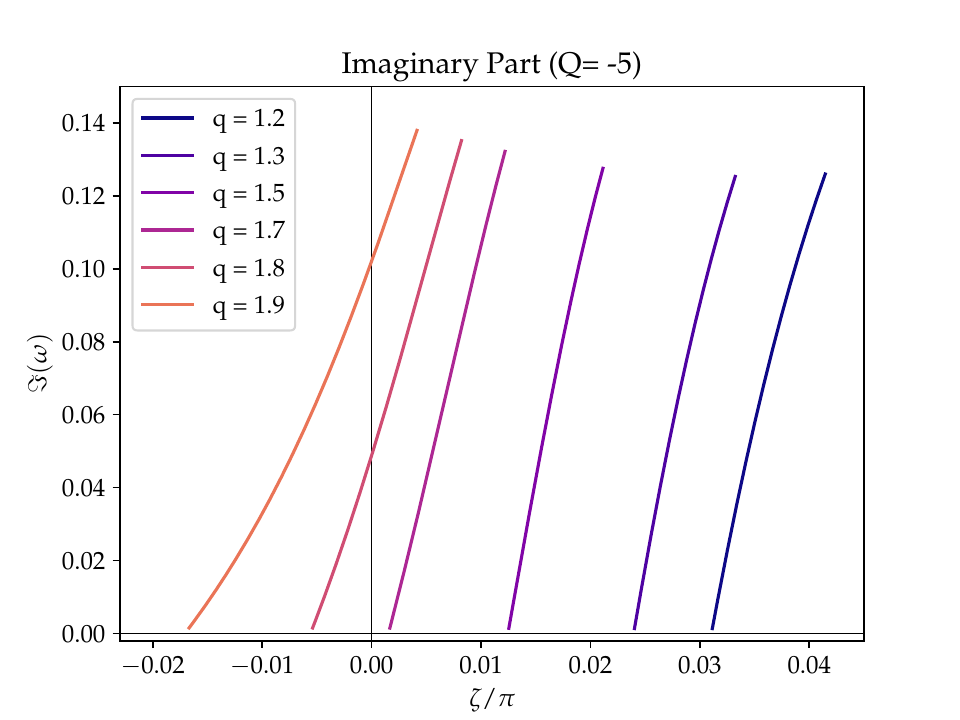}%
        \label{subfig:d31}%
    } 
\caption{Real (left) and imaginary (right) parts of the frequencies of superradiantly scattered charged scalar field modes as a function of the Robin parameter $\zeta $ (\ref{eq:Zeta}). 
We have fixed $m^{2}=-0.65$, $\ell =1$, $M=16$, $\Omega = 0.6$ and $k=1$. 
The black hole charge parameter is fixed to be $Q=-5$, and a selection of values of the scalar field charge $q$ are considered.}
\label{fig:frequencyQ-5}
\end{figure*}

Having provided evidence for superradiant scattering of a charged scalar field on a charged, rotating, BTZ black hole, we now consider the frequencies of the superradiant modes in order to determine whether or not a superradiant instability arises.

We begin, in Figure~\ref{fig:frequencyq0}, by considering the frequencies of neutral scalar field modes. 
The curves for a neutral black hole with $Q=0$ match those in \cite{Dappiaggi:2017pbe}, and those for $Q\neq 0$ have qualitatively similar properties (recall that for $q=0$ the radial equation (\ref{eq:radial}) does not depend on the sign of $Q$). 
The real parts of the frequencies decrease as the Robin parameter $\zeta $ increases.
It can be seen that the values of the frequencies do not change very much as the black hole charge $Q$ varies, but the Robin parameter does vary more significantly. This is in agreement with the energy fluxes shown in Figure~\ref{fig:Resultsq0}. 
The imaginary parts of the frequencies increase as the Robin parameter increases. 
All the superradiantly scattered neutral scalar field modes we consider here have positive imaginary part, and are therefore unstable, as in \cite{Dappiaggi:2017pbe} for $Q=0$. 
The range of values of the imaginary parts of the frequencies is not significantly changed by including a black hole charge $Q$, and therefore the timescale for the superradiant instability is also not significantly affected.

We now explore the frequencies of superradiantly scattered charged scalar field modes.
In Figures~\ref{fig:frequencyQ5}--\ref{fig:frequencyQ-5} we show the real and imaginary parts of the frequencies of the superradiantly scattered modes whose energy fluxes are depicted in Figure~\ref{fig:Branches}.
The features of the frequencies for the remaining superradiantly scattered modes shown in Figures~\ref{fig:Results}--\ref{fig:Rotation} are qualitatively similar so we relegate these to Appendix~\ref{sec:freqplots}.

Consider first $qQ>0$, so that the scalar field charge $q$ has the opposite sign to the black hole charge ${\widetilde{Q}}$ (\ref{eq:charge}). 
The frequencies in this case are shown in Figure~\ref{fig:frequencyQ5}.
In contrast to the neutral scalar case,  the real parts of the frequencies (left-hand plots in Figure~\ref{fig:frequencyQ5}) increase as the Robin parameter $\zeta $ increases, for all values of the black hole charge parameter $Q>0$, scalar field charge $q$ and rotation parameter $\Omega $ considered.

When $qQ>0$, the imaginary parts of the frequencies (right-hand plots in Figure~\ref{fig:frequencyQ5}) are negative for all parameter values examined, indicating that these modes are stable. 
This is a surprising result, since, for example, 
superradiantly scattered charged scalar field modes on a four-dimensional Reissner-Nordstr\"om-adS black hole are unstable \cite{Katagiri:2020mvm}.
We may understand this result heuristically as follows.

From (\ref{eq:flux1}, \ref{eq:currenthor}), an ingoing charged scalar field mode having $q>0$  and which is superradiantly scattered gives outgoing fluxes of both energy and charge at the event horizon. 
A superradiantly scattered charged scalar field mode is therefore extracting both energy and charge from the black hole.
At infinity, as in \cite{Katagiri:2020mvm}, the fluxes of both energy and charge vanish for all Robin boundary conditions.
When $qQ>0$, the black hole charge ${\widetilde {Q}}$ (\ref{eq:charge}) has the opposite sign to the scalar field charge $q$.
A scalar field mode with $q>0$, which is extracting charge from the black hole, therefore results in an increase in the magnitude of the black hole charge $|{\widetilde {Q}}|$  (since ${\widetilde {Q}}<0$ when $Q>0$). 
Since the magnitude of the black hole charge is increasing, the energy in the electromagnetic field is also increasing as a result of the superradiant scattering. 
In this case the energy extracted from the black hole by the superradiantly scattered charged scalar field mode is absorbed by the electromagnetic field rather than the scalar field.
Consequently, the scalar field does not grow with time and the superradiantly scattered mode is stable.

Turning now to $qQ<0$, the frequencies for superradiantly scattered modes in this case are shown in Figure~\ref{fig:frequencyQ-5}.
For the values of the scalar field charge $q>0$, black hole charge parameter $Q<0$ and rotation parameter $\Omega $ considered, the real parts of the frequencies (left-hand plots in Figure~\ref{fig:frequencyQ-5}) decrease as the Robin parameter $\zeta $ increases, as is the case for the neutral scalar field mode frequencies shown in Figure~\ref{fig:frequencyq0}.

When $qQ<0$, the imaginary parts of the frequencies (right-hand plots in Figure~\ref{fig:frequencyQ-5}) are positive for all parameter values studied, indicating that these superradiantly scattered modes are unstable, in contrast to the superradiantly scattered modes for $qQ>0$.
A heuristic argument for the instability of the superradiantly scattered charged scalar field modes with $qQ<0$ is as follows.

As discussed above, a superradiantly scattered ingoing charged scalar field mode with $q>0$ gives outgoing fluxes of both energy and charge, and therefore is extracting both energy and charge from the black hole. 
When $qQ<0$, the black hole charge ${\widetilde {Q}}$ (\ref{eq:charge}) has the same sign as the scalar field charge $q$. 
Therefore a scalar field mode with $q>0$ which is extracting charge from the black hole results in a decrease in the magnitude of the black hole charge $|{\widetilde {Q}}|$, and consequently a decrease in the energy in the electromagnetic field.
As a result, the energy extracted from the black hole is absorbed by the scalar field.  
Since the boundary conditions at infinity are reflecting, the scalar field wave is repeatedly superradiantly scattered off the black hole, resulting in an instability. 

For all values of the scalar field charge $q$, black hole charge parameter $Q$ and rotation parameter $\Omega $ studied in Figures~\ref{fig:frequencyQ5}--\ref{fig:frequencyQ-5}, we find that the imaginary part of the frequency of superradiantly scattered charged scalar field modes increases as the Robin parameter $\zeta $ increases.  
Therefore the stable superradiantly scattered modes for $qQ>0$ decay more slowly in time as $\zeta$ increases, while the unstable superradiantly scattered modes for $qQ<0$ are more rapidly growing with time as $\zeta $ increases.

\section{Conclusions}
\label{sec:conc}

In this paper we have explored the effect of black hole and scalar field charge on superradiant scattering on three-dim-ensional BTZ black holes.
We sought to compare superradiant scattering of a charged scalar field on a charged BTZ black hole with that on a four-dimensional  Reissner-Nordstr-\"om black hole. In particular, we examined whether charge superradiance is present on a nonrotating charged BTZ black hole. 
Superradiant amplification due to charge superradiant scattering on a Reissner-Nordstr\"om black hole is up to two orders of magnitude greater than that due to superradiant scattering on a rotating Kerr black hole, and we investigated whether similar enhancement also exists on BTZ black holes.

We considered separable mode solutions of the charged scalar field equation  on the charged generalization of the rotating BTZ black hole metric \cite{Martinez:1999qi}.
Working in the frequency domain, we find, as in the neutral scalar field case, that modes with real frequency do not exhibit superradiant scattering.
For modes with complex frequencies, following \cite{Teukolsky:1974yv,Dappiaggi:2017pbe}, we define superradiant scattering as occurring if the ingoing flux of energy due to an ingoing scalar field mode is negative (in other words, if an ingoing mode results in an outgoing flux of energy it is said to be superradiantly scattered).  
We find that it is necessary for the black hole to be rotating in order for superradiant scattering to occur.
Therefore, there is no superradiant scattering for nonrotating BTZ black holes, unlike the situation for four-dimensional, Reissner-Nordstr\"om  black holes.
Superradiantly scattered modes can only lie in a region of the complex frequency plane satisfying the inequality (\ref{eq:SRcriterion}). 

We use a simple numerical method, applicable to modes in the superradiant scattering regime, to provide evidence for the existence of superradiantly scattered charged scalar field modes when the black hole charge is nonzero.
We have not performed an exhaustive search of the parameter space, but instead considered a sample of black holes. For the region of parameter space explored, the presence of black hole and scalar field charges results in a flux of outgoing energy which is about an order of magnitude larger than in the uncharged case.
However, the dominant parameter affecting the magnitude of the outgoing energy flux seems to be the black hole rotation rather than the charges.

We have examined the range of boundary conditions satisfied by the superradiantly scattered modes at infinity.
These boundary conditions are labelled by the Robin parameter $\zeta $. 
For most fixed values of the black hole and scalar field charge that we studied, superradiantly scattered modes correspond to values of $\zeta $ lying in a narrow interval. 
For a large black hole charge parameter $|Q|=5$ (setting $\ell =1$), we have found some values of the scalar field charge $q\sim 2$ where the interval of values of $\zeta $ is considerably wider than in the generic case.
We have also found some values of $Q$ and $q$ for which there are superradiantly scattered modes satisfying either Dirichlet or Neumann boundary conditions, which are absent in the neutral scalar field case.
Superradiantly scattered modes satisfying Dirichlet boundary conditions have also been found for charged scalar perturbations of a Coulomb-like adS black hole in nonlinear electrodynamics in three dimensions \cite{Gonzalez:2021vwp}.

Finally, we have considered the real and imaginary parts of the frequencies of superradiantly scattered modes.
When the black hole and scalar field charges have the same sign (corresponding to $qQ<0$), we provide evidence that the superradiantly scattered modes are unstable. We also find superradiantly scattered modes when the scalar field and black hole charges have opposite signs ($qQ>0$). In this latter case we provide evidence that the superradiantly scattered modes, at least for the values of the parameters $q>0$ and $Q>0$ studied, are stable. 
For both signs of $qQ$, the superradiantly scattered modes are ingoing at the event horizon but extract energy from the black hole. When $qQ<0$, superradiantly scattered modes decrease the magnitude of the black hole charge and thereby the energy in the electromagnetic field also decreases. This means that the energy extracted from the black hole is absorbed by the scalar field, resulting in an instability. 
In contrast, for $qQ>0$, superradiantly scattered modes increase the magnitude of the black hole charge and thereby the energy in the electromagnetic field also increases. This means that the energy extracted from the black hole is absorbed by the electromagnetic field rather than the scalar field with no consequent scalar field instability. 

Our numerical method has limited us to exploring a comparatively small region of the parameter space. In particular, we find reliable numerical results only when at least one of the scalar field charge $q$ or black hole charge parameter $|Q|$ is comparatively large. 
We have also fixed the black hole mass parameter $M$ and azimuthal quantum number $k$,  as well as the scalar field mass $m$. 
Furthermore, we have restricted our attention to a charged scalar field minimally coupled to the spacetime curvature. 
With a more sophisticated numerical method (such as that employed in \cite{DalBoscoFontana:2023syy,Fontana:2023dix}), it would be interesting to probe the parameter space more widely.
In particular, we conjecture that there will be other superradiantly scattered charged scalar field modes that are unstable, as well as other unstable (but nonsuperradiantly scattered) modes. 
Testing this conjecture requires an alternative numerical method and hence is an avenue for future work.

In this paper we have studied a classical charged scalar field. 
A natural application of our work would be to consider a quantum charged scalar field.
The study of a massless, conformally coupled quantum scalar field on a neutral BTZ black hole is comparatively straightforward due to the construction of the BTZ metric by identifying points in 
adS space-time \cite{Banados:1992gq,Banados:1992wn}.
In particular, when either Dirichlet or Neumann boundary conditions are applied, the maximal symmetry of adS can be exploited to enable the computation of the renormalized expectation value of the stress-energy tensor using the method of images \cite{Steif:1993zv,Lifschytz:1993eb}, see also \cite{Kothawala:2008sm,Shiraishi:1993nu,Shiraishi:1993qnr}. 
This method is not applicable when Robin boundary conditions are applied as these break the maximal symmetry of the underlying adS geometry \cite{Barroso:2019cwp,Morley:2020ayr,Namasivayam:2022bky}.
The ground state Green's function for a neutral scalar field with Robin boundary conditions applied is constructed in~\cite{Bussola:2017wki} using a mode sum decomposition.  

It would be interesting to explore what effect the superradiantly scattered modes we have found in this paper have on the definition of quantum states for a charged scalar field on a charged BTZ black hole.
On four-dimensional asymptotically flat black holes, the presence of superradiantly scattered modes introduces subtleties in the construction of quantum states, both in the rotating \cite{Frolov:1989jh,Ottewill:2000qh,Duffy:2005mz,Casals:2005kr,Casals:2012es} and char-ged scenarios \cite{Balakumar:2022yvx}, and
one might anticipate similar challenges on a BTZ black hole.
Neutral scalar field modes on a neutral BTZ black hole are given by hypergeometric functions which simplifies the analysis \cite{Dappiaggi:2017pbe,Bussola:2017wki}.
For a charged scalar field on a charged BTZ background there appears to be no simple closed-form expression for the modes, which will complicate the construction.
We therefore postpone further consideration of the quantum charged scalar field to future work.


\begin{acknowledgements}
We thank Sam Dolan for helpful discussions and the anonymous referee whose recommendations have significantly improved the paper.
\end{acknowledgements}

\section*{Declarations}
The authors declare that they have no conflict of interest.
The work of E.W.~is supported by the Lancaster-Manchester-Sheffield Consortium for Fundamental Physics under STFC grant ST/T001038/1.
E.W.~also acknowledges the support of the Institut Henri Poincar\'{e} (UAR 839 CNRS-Sorbonne Universit\'{e}), and LabEx CARMIN (ANR-10-LABX-59-01). 
This research has also received funding from the European Union's Horizon 2020 research and innovation program under the H2020-MSCA-RISE-2017 Grant No.~FunFiCO-777740.
Data supporting this publication can be freely downloaded from the University of Sheffield Research Data Repository at {\url {https://doi.org/10.15131/shef.data.23717859}}, under the terms of the Creative Commons Attribution (CC--BY) licence. 
This version of the article has been accepted for publication, after peer review 
but is not the Version of Record and does not reflect post-acceptance improvements, or any
corrections. The Version of Record is available online at: {\url {http://dx.doi.org/10.1140/epjc/s10052-024-12910-7}}.

\appendix
\section{Numerical method}
\label{sec:method}

Our goal is to find complex frequencies $\omega $ such that the corresponding solution $X_{\omega \ell }(r)$ of the radial equation (\ref{eq:radial}) satisfies ingoing boundary conditions (\ref{eq:ingoing}) at the horizon and Robin boundary conditions (\ref{eq:XinfinityRobin}) at infinity. 
There are two key components of our method, integrating the radial equation for a given complex frequency $\omega $, and a root-finding algorithm to locate frequencies for which the boundary conditions at infinity are satisfied.

Given a complex frequency $\omega $, we impose ingoing boundary conditions (\ref{eq:ingoing}) on the radial function $X_{\omega k}(r)$ at $r=r_{h}+\epsilon $, where $\epsilon \ll 1$. 
For $r\gg r_{h}$, the function $X_{\omega k}(r)$ takes the form (\ref{eq:Xinfinity1}) for complex constants $C_{\omega k}$, $D_{\omega \ell }$. 
We rewrite the radial equation (\ref{eq:radial}) in terms of the radial coordinate $r$ and a new dependent variable $Y_{\omega k}(r)=r^{\frac{1}{2}(1-{\sqrt {1+4\mu ^{2}}})}$ $X_{\omega k}(r)$.
We numerically integrate this new radial equation from $r=r_{h}+\epsilon $ to $r=r_{{\text{max}}}$, where $r_{{\text{max}}}\gg r_{h}$, using  {\tt {MATHEMATICA}}'s built-in {\tt {NDSolve}} command.
We have tested the numerical integration by using a sample of both stable and unstable QNM frequencies for a charged scalar field on nonrotating charged BTZ black hole given in \cite{Fontana:2023dix}.
The value of $D_{\omega k}$ can be found as the limit of $Y_{\omega k}(r)$ as $r\rightarrow \infty $.
The value of $C_{\omega k}$ is found from the limit of $r^{1+{\sqrt {1+4\mu ^{2}}}}Y_{\omega k}'(r)$ as $r\rightarrow \infty $. 
We require the numerical integration of the radial equation (\ref{eq:radial}) to very high precision in order to extract the constants $C_{\omega k}$ and $D_{\omega k}$ to a reasonable accuracy.

For a general frequency $\omega $, the constants $C_{\omega k}$ and $D_{\omega k}$ thus found will be complex. 
To apply Robin boundary conditions (\ref{eq:XinfinityRobin}), we require the ratio $D_{\omega k}/C_{\omega k}$ to be real.
For fixed $\Re (\omega )$ in the interval $0<\Re (\omega )<k\Omega /\ell $ for which superradiant scattering is possible, we use {\tt {MATHEMATICA}}'s inbuilt root-finding command {\tt {FindRoot}} to find the value of $\Im (\omega )$ for which the imaginary part of $D_{\omega k}/C_{\omega k}$ vanishes.

From \cite{Dappiaggi:2017pbe}, the spectrum of QNM frequencies for the neutral scalar field on a neutral BTZ black hole is very complicated, with many modes located very close together in $(\omega, \zeta )$ space (where $\zeta $ is the parameter governing the Robin boundary conditions (\ref{eq:Zeta})).
We suspect that the same is true for a charged scalar field on a charged BTZ black hole.
This presents challenges for the root-finding algorithm, which can jump from one branch of QNM frequencies to another as $\Re(\omega )$ is varied. 

When $Q=0=q$, the root-finding algorithm reproduces the results presented in \cite{Dappiaggi:2017pbe} (see the $Q=0$ curves in Figures \ref{fig:Resultsq0} and \ref{fig:frequencyq0}). 
This is a highly nontrivial check, since we are directly integrating the radial equation (\ref{eq:radial}) while the authors of \cite{Dappiaggi:2017pbe} exploit the fact that the radial functions are given in terms of hypergeometric functions.

For $q\neq 0$, the root-finding algorithm gives robust results for sufficiently large $q$ and $|Q|$, as presented in Section \ref{sec:rotating} and Appendix \ref{sec:freqplots}. 
We have verified that changing the parameters $\epsilon $ and $r_{{\text{max}}}$ in the numerical integration procedure yields plots which are indistinguishable from those presented in Section~\ref{sec:rotating}  and Appendix \ref{sec:freqplots}, and estimate the maximum relative error in the QNM frequencies to be of the order of $\mathcal{O}(10^{-5})$.

\section{QNM frequencies}
\label{sec:freqplots}

\begin{figure*}[ht!]
    \subfloat[]{%
        \includegraphics[width=.50\linewidth]{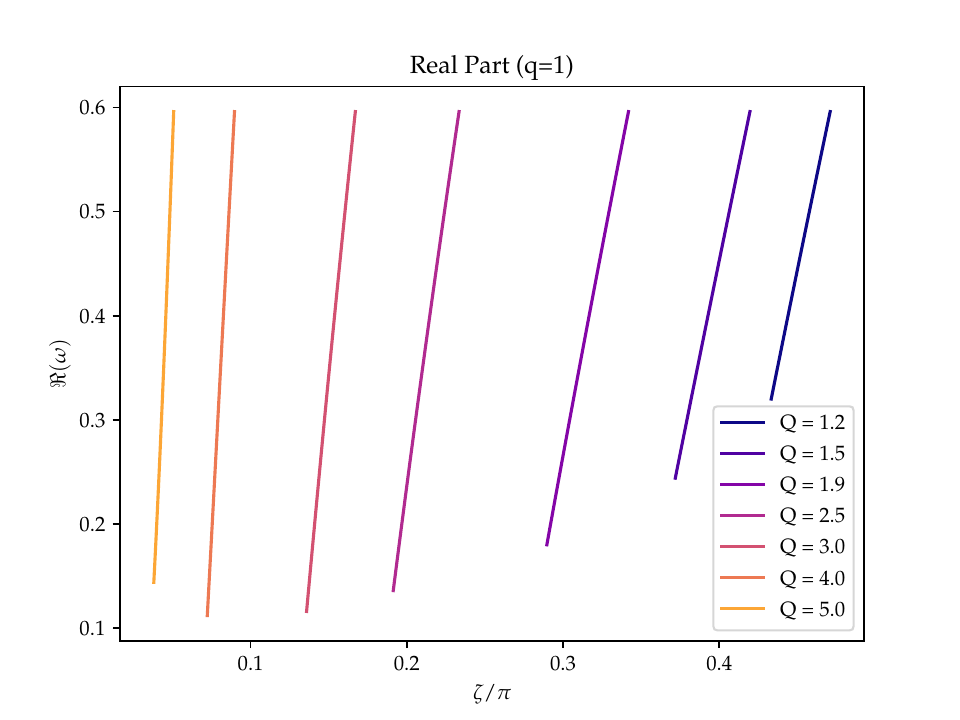}%
        \label{subfig:a1}%
    }\hfill
    \subfloat[]{%
        \includegraphics[width=.50\linewidth]{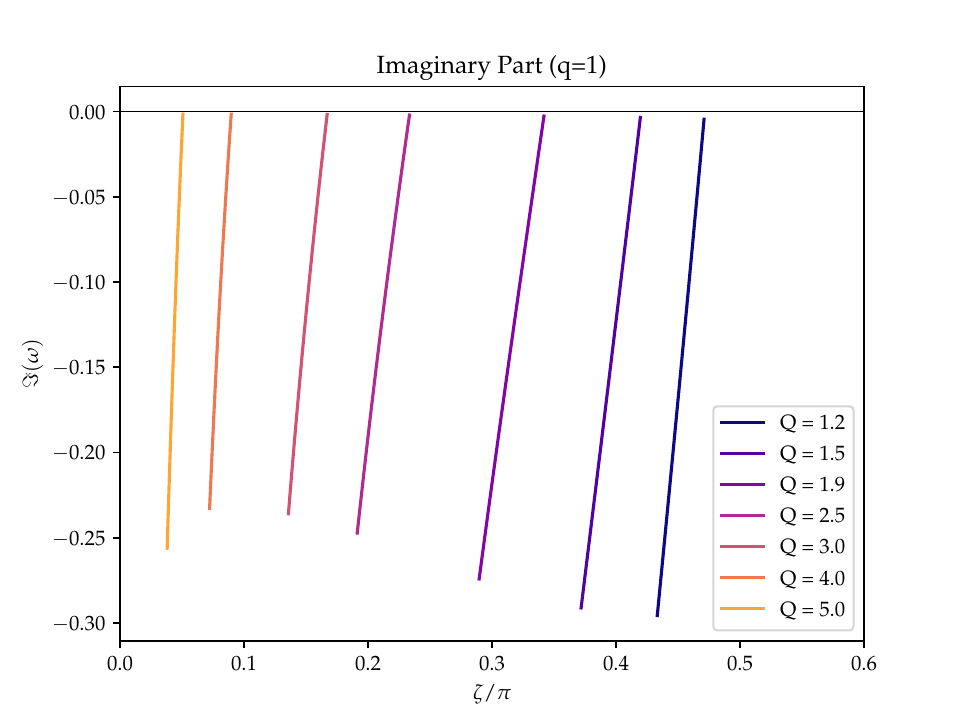}%
        \label{subfig:b1}%
    } 
\\
\subfloat[]{%
        \includegraphics[width=.50\linewidth]{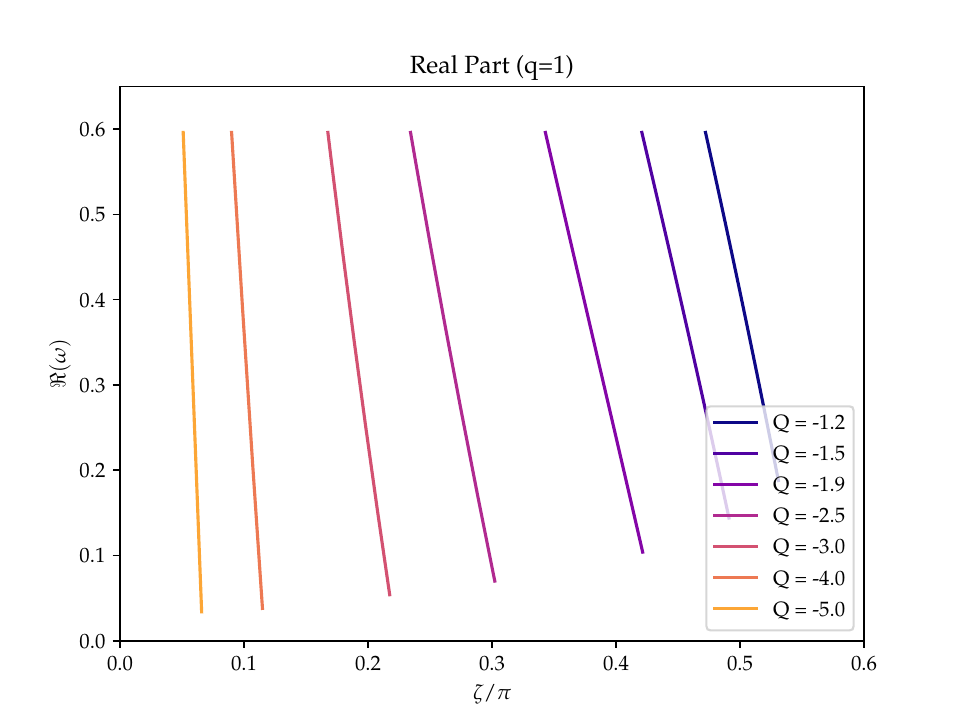}%
        \label{subfig:c1}%
    }\hfill
    \subfloat[]{%
        \includegraphics[width=.50\linewidth]{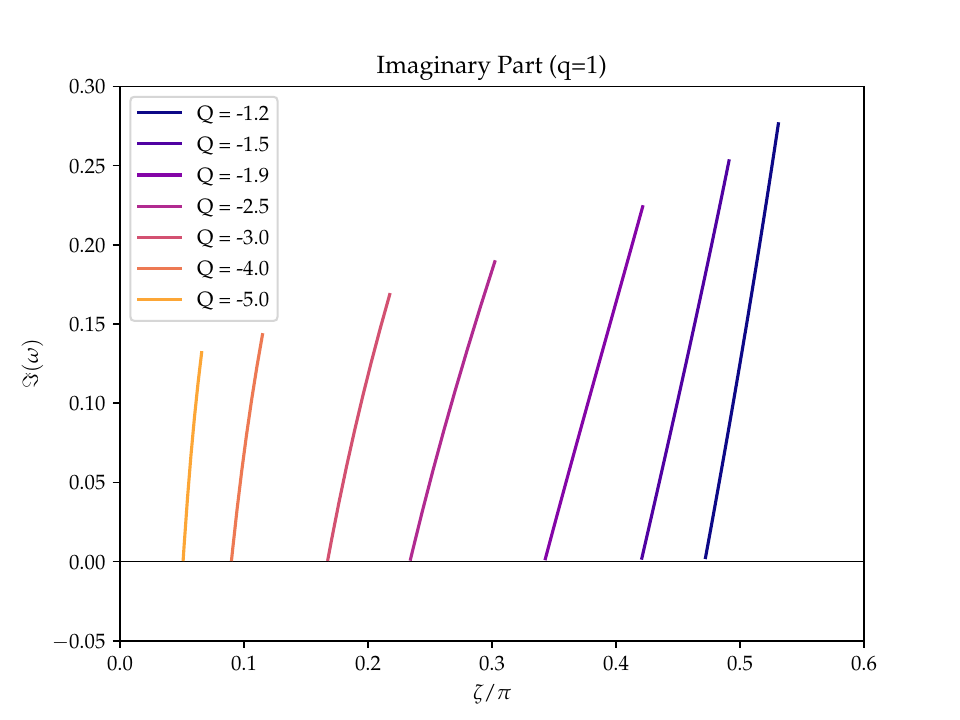}%
        \label{subfig:d1}%
    } 
\caption{Real (left) and imaginary (right) parts of the frequencies of superradiantly scattered charged scalar field modes as a function of the Robin parameter $\zeta $ (\ref{eq:Zeta}). 
We have fixed $m^{2}=-0.65$, $\ell =1$, $M=16$, $\Omega = 0.6$ and $k=1$. 
The values of the black hole charge parameter $Q$ and scalar field charge $q$ are as given in the legends.}
\label{fig:frequency}
\end{figure*}

\begin{figure*}[ht!]
    \subfloat[]{%
        \includegraphics[width=.50\linewidth]{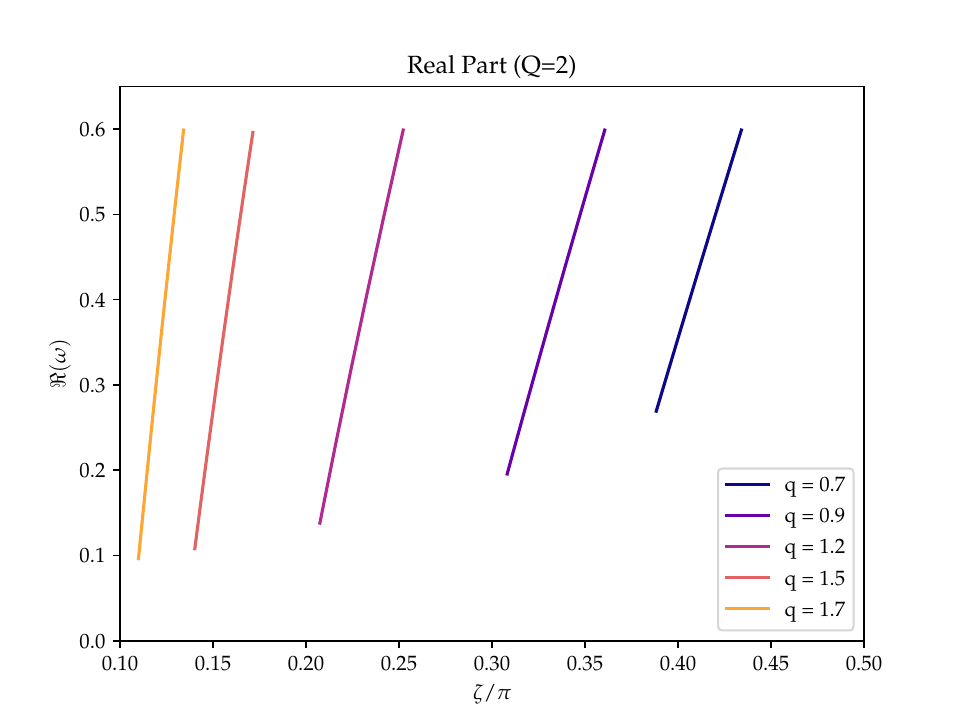}%
        \label{subfig:a11}%
    }\hfill
    \subfloat[]{%
        \includegraphics[width=.50\linewidth]{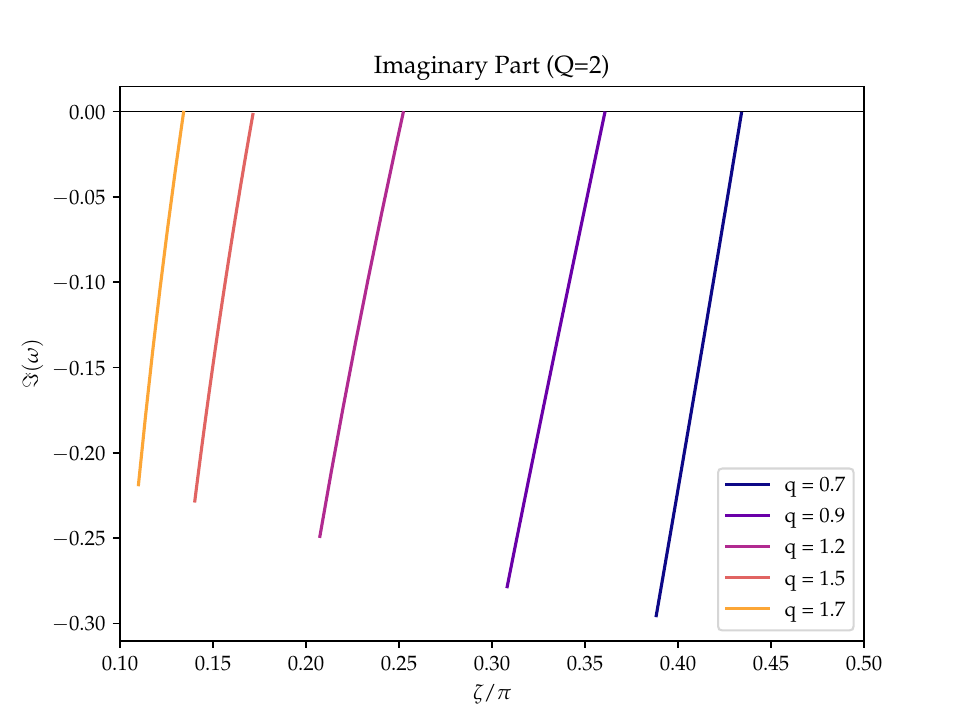}%
        \label{subfig:b11}%
    } 
\\
\subfloat[]{%
        \includegraphics[width=.50\linewidth]{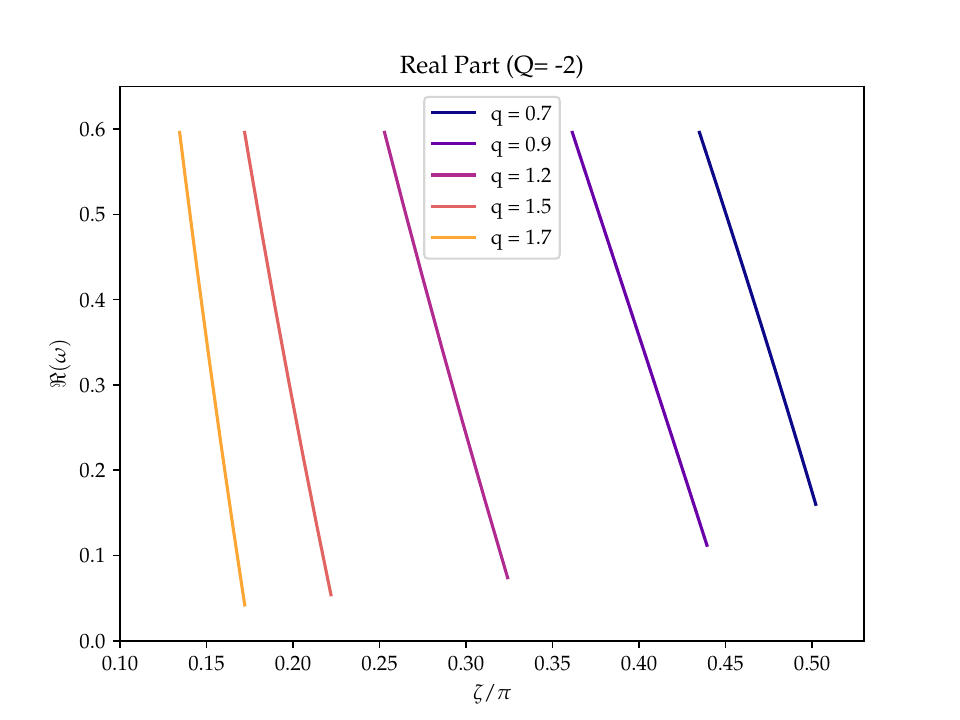}%
        \label{subfig:c11}%
    }\hfill
    \subfloat[]{%
        \includegraphics[width=.50\linewidth]{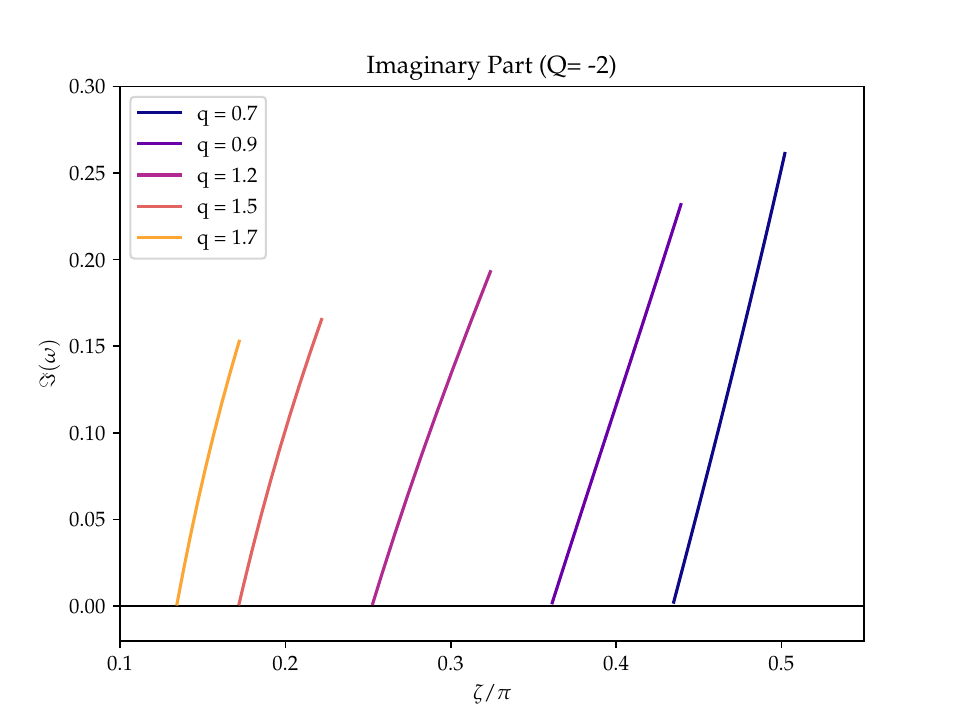}%
        \label{subfig:d11}%
    } 
\caption{Real (left) and imaginary (right) parts of the frequencies of superradiantly scattered charged scalar field modes as a function of the Robin parameter $\zeta $ (\ref{eq:Zeta}). 
We have fixed $m^{2}=-0.65$, $\ell =1$, $M=16$, $\Omega = 0.6$ and $k=1$. 
The values of the black hole charge parameter $Q$ and scalar field charge $q$ are as given in the legends.}
\label{fig:frequency1}
\end{figure*}

\begin{figure*}[ht!]
    \subfloat[]{%
        \includegraphics[width=.50\linewidth]{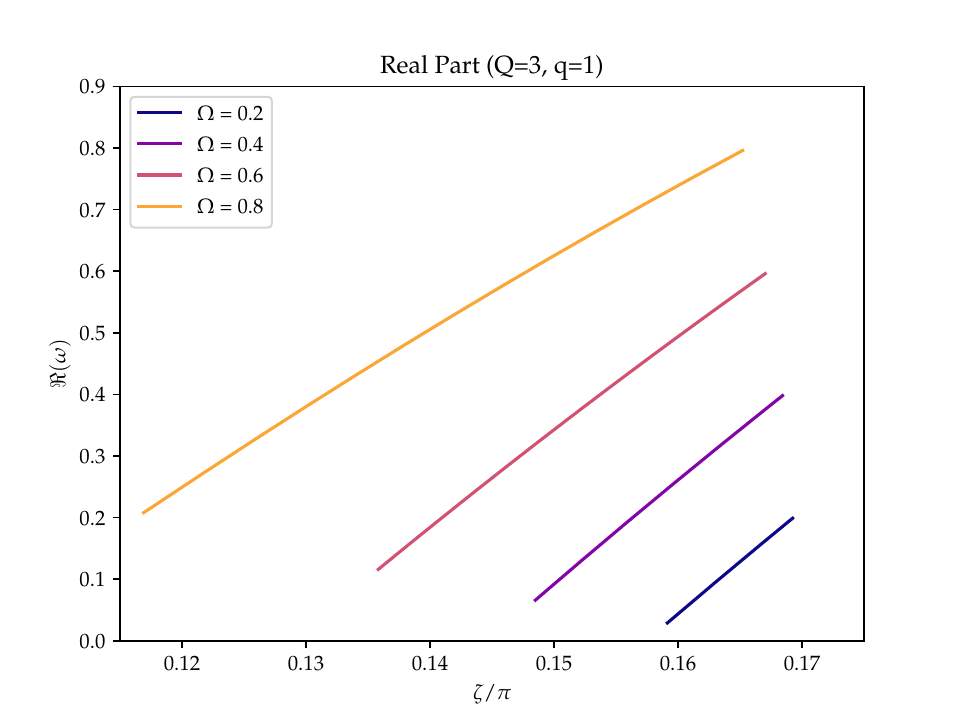}%
        \label{subfig:e1}%
    }\hfill
    \subfloat[]{%
        \includegraphics[width=.50\linewidth]{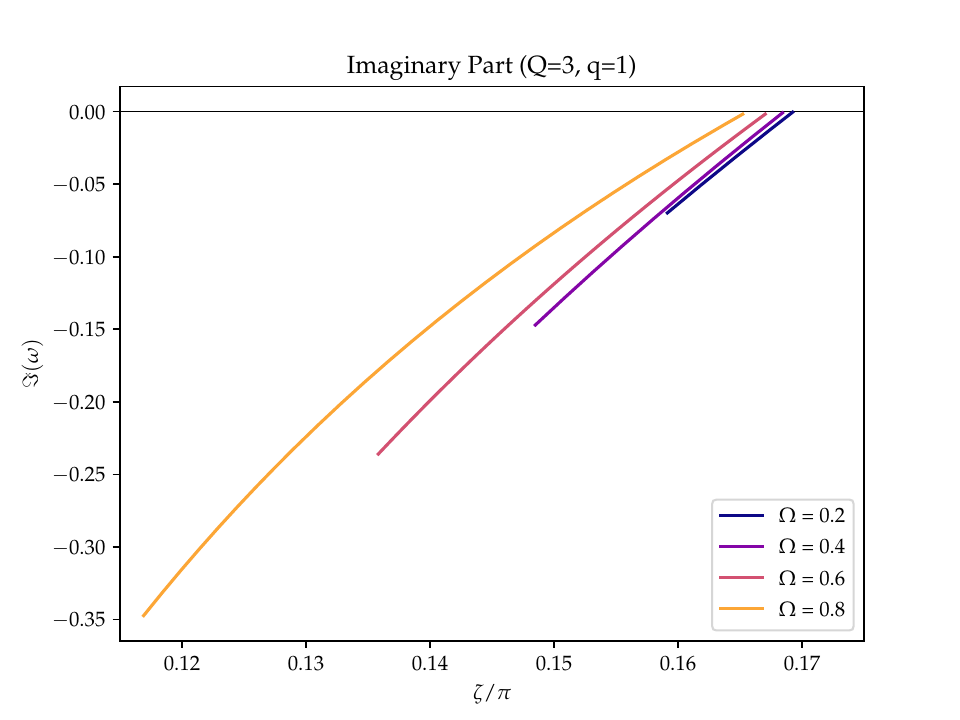}%
        \label{subfig:f1}%
    } \\
    \subfloat[]{%
        \includegraphics[width=.50\linewidth]{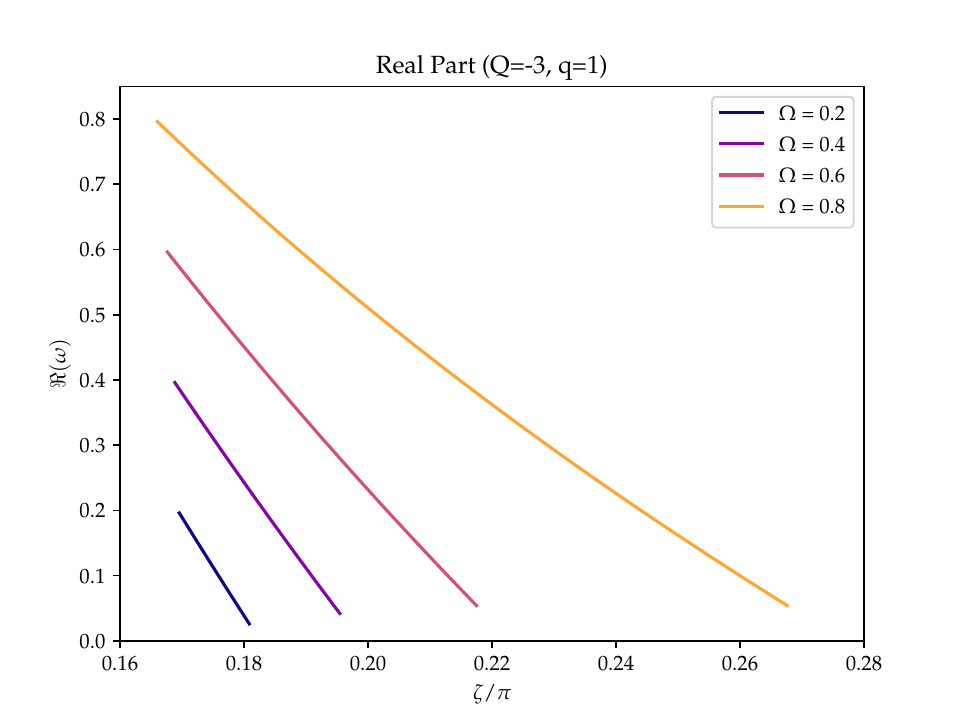}%
        \label{subfig:e11}%
    }\hfill
    \subfloat[]{%
        \includegraphics[width=.50\linewidth]{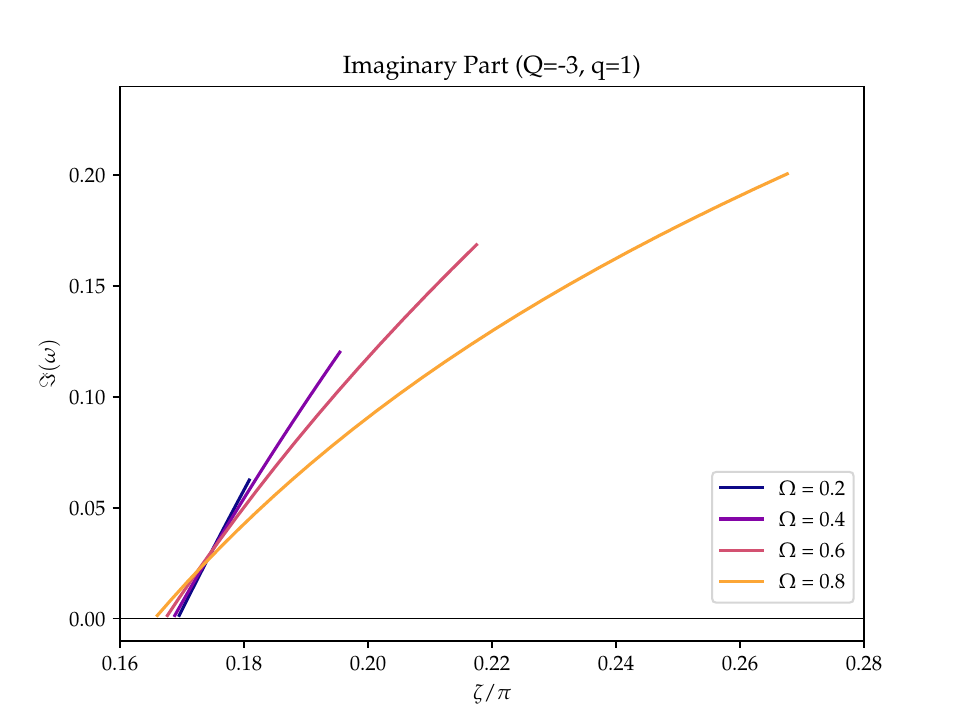}%
        \label{subfig:f11}%
    } 
\caption{Real (left) and imaginary (right) parts of the frequencies of superradiantly scattered charged scalar field modes as a function of the Robin parameter $\zeta $ (\ref{eq:Zeta}). 
We have fixed $m^{2}=-0.65$, $\ell =1$, $M=16$, $Q=\pm3$, $q=1$ and $k=1$. 
The values of the rotation parameter $\Omega $ are as given in the legends.}
\label{fig:frequencyrot}
\end{figure*}

In Figures~\ref{fig:frequency}--\ref{fig:frequencyrot} we present the real and imaginary parts of the frequencies of superradiantly scattered charged scalar field modes whose energy fluxes are depicted in Figures~\ref{fig:Results}--\ref{fig:Rotation}.
The qualitative behaviour of these frequencies is very similar to those shown in Figures~\ref{fig:frequencyQ5}--\ref{fig:frequencyQ-5}.
When $qQ>0$ and the scalar field charge $q$ has the opposite sign to the black hole charge ${\widetilde{Q}}$ (\ref{eq:charge}), the real parts of the frequencies are increasing as the Robin parameter $\zeta $ (\ref{eq:Zeta}) increases.
When $qQ<0$ and the scalar field charge $q$ has the same sign as the black hole charge ${\widetilde{Q}}$ (\ref{eq:charge}), the real parts of the frequencies are decreasing as the Robin parameter $\zeta $ (\ref{eq:Zeta}) increases.
For all the superradiantly scattered modes considered here, we find that the imaginary parts of the frequencies are negative when $qQ>0$ and positive when $qQ<0$.
As discussed in Section \ref{sec:frequencies}, we deduce that superradiantly scattered charged scalar field modes are unstable when the scalar field has a charge with the same sign as that of the black hole, but if the scalar field charge has the opposite sign to that of the black hole, the superradiantly scattered modes considered here are stable.
When $qQ>0$, the imaginary parts of the frequencies are increasing as the Robin paramter $\zeta $ increases, implying that the rate at which these modes decay in time decreases as $\zeta $ increases.
For $qQ<0$, it can be seen that the imaginary parts of the frequencies are again increasing as $\zeta $ increases, and thus the superradiantly scattered modes grow more rapidly with increasing $\zeta $.


\begin{thebibliography}{10}
\providecommand{\url}[1]{{#1}}
\providecommand{\urlprefix}{URL }
\expandafter\ifx\csname urlstyle\endcsname\relax
  \providecommand{\doi}[1]{DOI \discretionary{}{}{}#1}\else
  \providecommand{\doi}{DOI \discretionary{}{}{}\begingroup
  \urlstyle{rm}\Url}\fi

\bibitem{Brito:2015oca}
R.~Brito, V.~Cardoso, P.~Pani, Lect. Notes Phys. \textbf{906}, (2015).
\newblock \doi{10.1007/978-3-319-19000-6}

\bibitem{Misner:1972kx}
C.W. Misner, Phys. Rev. Lett. \textbf{28}, 994 (1972).
\newblock \doi{10.1103/PhysRevLett.28.994}

\bibitem{Press:1972zz}
W.H. Press and S.A. Teukolsky, Nature \textbf{238}, 211 (1972).
\newblock \doi{10.1038/238211a0}

\bibitem{Chandrasekhar:1985kt}
S.~Chandrasekhar, \emph{{The mathematical theory of black holes}} (Oxford
  University Press, 1985).

\bibitem{Teukolsky:1974yv}
S.~A.~Teukolsky, W.~H.~Press,
Astrophys. J. \textbf{193}, 443 (1974).
\newblock \doi{10.1086/153180} 

\bibitem{Penrose:1971uk}
R.~Penrose, R.M. Floyd, Nature \textbf{229}, 177 (1971).
\newblock \doi{10.1038/physci229177a0}

\bibitem{Bekenstein:1973mi}
J.D. Bekenstein, Phys. Rev. D \textbf{7}, 949 (1973).
\newblock \doi{10.1103/PhysRevD.7.949}

\bibitem{Nakamura:1976nc}
T.~Nakamura, H.~Sato, Phys. Lett. B \textbf{61}, 371 (1976).
\newblock \doi{10.1016/0370-2693(76)90591-8}

\bibitem{DiMenza:2014vpa}
L.~Di~Menza, J.P. Nicolas, Class. Quant. Grav. \textbf{32}, 145013 (2015).
\newblock \doi{10.1088/0264-9381/32/14/145013}

\bibitem{Benone:2015bst}
C.L. Benone, L.C.B. Crispino, Phys. Rev. D \textbf{93}, 024028 (2016).
\newblock \doi{10.1103/PhysRevD.93.024028}

\bibitem{DiMenza:2019zli}
L.~Di~Menza, J.p. Nicolas, M.~Pellen, Gen. Rel. Grav. \textbf{52}, 8 (2020).
\newblock \doi{10.1007/s10714-020-2656-5}

\bibitem{Detweiler:1980uk}
S.L.~Detweiler,
Phys. Rev. D \textbf{22}, 2323 (1980).
\newblock \doi{10.1103/PhysRevD.22.2323}

\bibitem{Zouros:1979iw}
T.J.M.~Zouros, D.M.~Eardley,
Annals Phys. \textbf{118}, 139 (1979).
\newblock \doi{10.1016/0003-4916(79)90237-9}

\bibitem{Dolan:2007mj}
S.R.~Dolan,
Phys. Rev. D \textbf{76}, 084001 (2007).
\newblock \doi{10.1103/PhysRevD.76.084001}

\bibitem{Okawa:2014nda}
H.~Okawa, H.~Witek, V.~Cardoso,
Phys. Rev. D \textbf{89}, 104032 (2014).
\newblock \doi{10.1103/PhysRevD.89.104032}

\bibitem{East:2013mfa}
W.E.~East, F.M.~Ramazano\u{g}lu, F.~Pretorius,
Phys. Rev. D \textbf{89}, 061503 (2014).
\newblock \doi{10.1103/PhysRevD.89.061503}

\bibitem{Hod:2013nn}
S.~Hod,
Phys. Lett. B \textbf{718}, 1489 (2013).
\newblock \doi{10.1016/j.physletb.2012.12.013}

\bibitem{Hod:2015hza}
S.~Hod,
Phys. Rev. D \textbf{91}, 044047 (2015).
\newblock
\doi{10.1103/PhysRevD.91.044047}

\bibitem{Herdeiro:2013pia}
C.A.R.~Herdeiro, J.C.~Degollado, H.F.~R\'unarsson,
Phys. Rev. D \textbf{88}, 063003 (2013).
\newblock \doi{10.1103/PhysRevD.88.063003}

\bibitem{Degollado:2013bha}
J.C.~Degollado and C.A.R.~Herdeiro,
Phys. Rev. D \textbf{89}, 063005 (2014).
\newblock \doi{10.1103/PhysRevD.89.063005}

\bibitem{Sanchis-Gual:2015lje}
N.~Sanchis-Gual, J.C.~Degollado, P.J.~Montero, J.A.~Font, C.~Herdeiro,
Phys. Rev. Lett. \textbf{116}, 141101 (2016).
\newblock \doi{10.1103/PhysRevLett.116.141101}

\bibitem{Sanchis-Gual:2016tcm}
N.~Sanchis-Gual, J.C.~Degollado, C.~Herdeiro, J.A.~Font, P.J.~Montero,
Phys. Rev. D \textbf{94}, 044061 (2016).
\newblock \doi{10.1103/PhysRevD.94.044061}

\bibitem{Cardoso:2004hs}
V.~Cardoso, O.J.C. Dias, Phys. Rev. D \textbf{70}, 084011 (2004).
\newblock \doi{10.1103/PhysRevD.70.084011}

\bibitem{Cardoso:2006wa}
V.~Cardoso, O.J.C. Dias, S.~Yoshida, Phys. Rev. D \textbf{74}, 044008 (2006).
\newblock \doi{10.1103/PhysRevD.74.044008}

\bibitem{Kunduri:2006qa}
H.K. Kunduri, J.~Lucietti, H.S. Reall, Phys. Rev. D \textbf{74}, 084021 (2006).
\newblock \doi{10.1103/PhysRevD.74.084021}

\bibitem{Kodama:2007sf}
H.~Kodama, Prog. Theor. Phys. Suppl. \textbf{172}, 11 (2008).
\newblock \doi{10.1143/PTPS.172.11}

\bibitem{Gubser:2008px}
S.S. Gubser, Phys. Rev. D \textbf{78}, 065034 (2008).
\newblock \doi{10.1103/PhysRevD.78.065034}

\bibitem{Aliev:2008yk}
A.N. Aliev, O.~Delice, Phys. Rev. D \textbf{79}, 024013 (2009).
\newblock \doi{10.1103/PhysRevD.79.024013}

\bibitem{Murata:2008xr}
K.~Murata, Prog. Theor. Phys. \textbf{121}, 1099 (2009).
\newblock \doi{10.1143/PTP.121.1099}

\bibitem{Kodama:2009rq}
H.~Kodama, R.A. Konoplya, A.~Zhidenko, Phys. Rev. D \textbf{79}, 044003 (2009).
\newblock \doi{10.1103/PhysRevD.79.044003}

\bibitem{Uchikata:2009zz}
N.~Uchikata, S.~Yoshida, T.~Futamase, Phys. Rev. D \textbf{80}, 084020 (2009).
\newblock \doi{10.1103/PhysRevD.80.084020}

\bibitem{Dias:2011tj}
O.J.C. Dias, P.~Figueras, S.~Minwalla, P.~Mitra, R.~Monteiro, J.E. Santos, JHEP
  \textbf{08}, 117 (2012).
\newblock \doi{10.1007/JHEP08(2012)117}

\bibitem{Li:2012rx}
R.~Li, Phys. Lett. B \textbf{714}, 337 (2012).
\newblock \doi{10.1016/j.physletb.2012.07.015}

\bibitem{Cardoso:2013pza}
V.~Cardoso, O.J.C. Dias, G.S. Hartnett, L.~Lehner, J.E. Santos, JHEP
  \textbf{04}, 183 (2014).
\newblock \doi{10.1007/JHEP04(2014)183}

\bibitem{Wang:2014eha}
M.~Wang, C.~Herdeiro, Phys. Rev. D \textbf{89}, 084062 (2014).
\newblock \doi{10.1103/PhysRevD.89.084062}

\bibitem{Wang:2015fgp}
M.~Wang, C.~Herdeiro, Phys. Rev. D \textbf{93}, 064066 (2016).
\newblock \doi{10.1103/PhysRevD.93.064066}

\bibitem{Delice:2015zga}
O.~Delice, T.~Dur\u{g}ut, Phys. Rev. D \textbf{92}, 024053 (2015).
\newblock \doi{10.1103/PhysRevD.92.024053}

\bibitem{Green:2015kur}
S.R. Green, S.~Hollands, A.~Ishibashi, R.M. Wald, Class. Quant. Grav.
  \textbf{33}, 125022 (2016).
\newblock \doi{10.1088/0264-9381/33/12/125022}

\bibitem{Bosch:2016vcp}
P.~Bosch, S.R. Green, L.~Lehner, Phys. Rev. Lett. \textbf{116}, 141102
  (2016).
\newblock \doi{10.1103/PhysRevLett.116.141102}

\bibitem{Dias:2016pma}
O.J.C. Dias, R.~Masachs, JHEP \textbf{02}, 128 (2017).
\newblock \doi{10.1007/JHEP02(2017)128}

\bibitem{BarraganAmado:2018zpa}
J.~Barrag\'an~Amado, B.~Carneiro Da~Cunha, E.~Pallante, Phys. Rev. D
  \textbf{99}, 105006 (2019).
\newblock \doi{10.1103/PhysRevD.99.105006}

\bibitem{Chesler:2018txn}
P.M. Chesler, D.A. Lowe, Phys. Rev. Lett. \textbf{122}, 181101 (2019).
\newblock \doi{10.1103/PhysRevLett.122.181101}

\bibitem{Li:2019tns}
R.~Li, Y.~Zhao, T.~Zi, X.~Chen, Phys. Rev. D \textbf{99}, 084045 (2019).
\newblock \doi{10.1103/PhysRevD.99.084045}

\bibitem{Katagiri:2020mvm}
T.~Katagiri, T.~Harada, Class. Quant. Grav. \textbf{38}, 135026 (2021).
\newblock \doi{10.1088/1361-6382/abfed6}

\bibitem{Chesler:2021ehz}
P.M. Chesler, Phys. Rev. D \textbf{105}, 024026 (2022).
\newblock \doi{10.1103/PhysRevD.105.024026}

\bibitem{Ishii:2021ncf}
T.~Ishii, SciPost Phys. Proc. \textbf{4}, 008 (2021).
\newblock \doi{10.21468/SciPostPhysProc.4.008}

\bibitem{Ishii:2022lwc}
T.~Ishii, Y.~Kaku, K.~Murata, JHEP \textbf{10}, 024 (2022).
\newblock \doi{10.1007/JHEP10(2022)024}

\bibitem{Banados:1992gq}
M.~Banados, M.~Henneaux, C.~Teitelboim, J.~Zanelli, Phys. Rev. D \textbf{48},
  1506 (1993).
\newblock \doi{10.1103/PhysRevD.48.1506}.
\newblock [Erratum: Phys. Rev. D \textbf{88}, 069902 (2013)]

\bibitem{Banados:1992wn}
M.~Banados, C.~Teitelboim, J.~Zanelli, Phys. Rev. Lett. \textbf{69}, 1849
  (1992).
\newblock \doi{10.1103/PhysRevLett.69.1849}

\bibitem{Carlip:1995qv}
S.~Carlip, Class. Quant. Grav. \textbf{12}, 2853 (1995).
\newblock \doi{10.1088/0264-9381/12/12/005}

\bibitem{Ichinose:1994rg}
I.~Ichinose, Y.~Satoh,
Nucl. Phys. B \textbf{447}, 340 (1995).
\newblock \doi{10.1016/0550-3213(95)00197-Z}

\bibitem{Ghoroku:1994np}
K.~Ghoroku, A.L.~Larsen,
Phys. Lett. B \textbf{328}, 28 (1994).
\newblock \doi{10.1016/0370-2693(94)90423-5} 

\bibitem{Ortiz:2011wd}
L.~Ortiz, Phys. Rev. D \textbf{86}, 047703 (2012).
\newblock \doi{10.1103/PhysRevD.86.047703}

\bibitem{Ortiz:2018ddt}
L.~Ort\'\i{}z, N.~Bret\'on, Mod. Phys. Lett. A \textbf{34}, 1950251 (2019).
\newblock \doi{10.1142/S0217732319502511}

\bibitem{Ishibashi:2004wx}
A.~Ishibashi, R.M. Wald, Class. Quant. Grav. \textbf{21}, 2981 (2004).
\newblock \doi{10.1088/0264-9381/21/12/012}

\bibitem{Dappiaggi:2017pbe}
C.~Dappiaggi, H.R.C. Ferreira, C.A.R. Herdeiro, Phys. Lett. B \textbf{778}, 146
  (2018).
\newblock \doi{10.1016/j.physletb.2018.01.018}

\bibitem{Ferreira:2017tnc}
H.R.C. Ferreira, C.A.R. Herdeiro, Phys. Rev. D \textbf{97}, 084003 (2018).
\newblock \doi{10.1103/PhysRevD.97.084003}

\bibitem{Garcia:1999py}
A.~Garcia,   (1999).
\newblock \doi{10.48550/arXiv.hep-th/9909111}

\bibitem{Clement:1993kc}
G.~Clement, Class. Quant. Grav. \textbf{10}, L49 (1993).
\newblock \doi{10.1088/0264-9381/10/5/002}

\bibitem{Clement:1995zt}
G.~Clement, Phys. Lett. B \textbf{367}, 70 (1996).
\newblock \doi{10.1016/0370-2693(95)01464-0}

\bibitem{Kamata:1995zu}
M.~Kamata, T.~Koikawa, Phys. Lett. B \textbf{353}, 196 (1995).
\newblock \doi{10.1016/0370-2693(95)00583-7}

\bibitem{Cataldo:1999fh}
M.~Cataldo, P.~Salgado, Phys. Lett. B \textbf{448}, 20 (1999).
\newblock \doi{10.1016/S0370-2693(99)00035-0}

\bibitem{Martinez:1999qi}
C.~Martinez, C.~Teitelboim, J.~Zanelli, Phys. Rev. D \textbf{61}, 104013
  (2000).
\newblock \doi{10.1103/PhysRevD.61.104013}

\bibitem{Tang:2016vmu}
Z.Y. Tang, C.Y. Zhang, M.~Kord~Zangeneh, B.~Wang, J.~Saavedra, Eur. Phys. J. C
  \textbf{77}, 390 (2017).
\newblock \doi{10.1140/epjc/s10052-017-4966-7}

\bibitem{Bussola:2017wki}
F.~Bussola, C.~Dappiaggi, H.R.C. Ferreira, I.~Khavkine, Phys. Rev. D
  \textbf{96}, 105016 (2017).
\newblock \doi{10.1103/PhysRevD.96.105016}

\bibitem{Garbarz:2017wzv}
A.~Garbarz, J.~La~Madrid, M.~Leston, Eur. Phys. J. C \textbf{77}, 807
  (2017).
\newblock \doi{10.1140/epjc/s10052-017-5385-5}

\bibitem{Breitenlohner:1982bm}
P.~Breitenlohner, D.Z. Freedman, Phys. Lett. B \textbf{115}, 197 (1982).
\newblock \doi{10.1016/0370-2693(82)90643-8}

\bibitem{Breitenlohner:1982jf}
P.~Breitenlohner, D.Z. Freedman, Annals Phys. \textbf{144}, 249 (1982).
\newblock \doi{10.1016/0003-4916(82)90116-6}

\bibitem{Kokkotas:1999bd}
K.D. Kokkotas, B.G. Schmidt, Living Rev. Rel. \textbf{2}, 2 (1999).
\newblock \doi{10.12942/lrr-1999-2}

\bibitem{Berti:2009kk}
E.~Berti, V.~Cardoso, A.O. Starinets, Class. Quant. Grav. \textbf{26}, 163001
  (2009).
\newblock \doi{10.1088/0264-9381/26/16/163001}

\bibitem{Konoplya:2011qq}
R.A. Konoplya, A.~Zhidenko, Rev. Mod. Phys. \textbf{83}, 793 (2011).
\newblock \doi{10.1103/RevModPhys.83.793}

\bibitem{Cho:2011sf}
H.T. Cho, A.S. Cornell, J.~Doukas, T.R. Huang, W.~Naylor, Adv. Math. Phys.
  \textbf{2012}, 281705 (2012).
\newblock \doi{10.1155/2012/281705}

\bibitem{Birmingham:2001hc}
D.~Birmingham,
Phys. Rev. D \textbf{64}, 064024, (2001).
\newblock \doi{10.1103/PhysRevD.64.064024}

\bibitem{Cardoso:2001hn}
V.~Cardoso, J.P.S. Lemos, Phys. Rev. D \textbf{63}, 124015 (2001).
\newblock \doi{10.1103/PhysRevD.63.124015}

\bibitem{Crisostomo:2004hj}
J.~Cris\'ostomo, S.~Lepe, J.~Saavedra, Class. Quant. Grav. \textbf{21}, 2801
  (2004).
\newblock \doi{10.1088/0264-9381/21/12/002}

\bibitem{Decanini:2009dn}
Y.~Decanini, A.~Folacci, Phys. Rev. D \textbf{79}, 044021 (2009).
\newblock \doi{10.1103/PhysRevD.79.044021}

\bibitem{Zhang:2011ec}
H.b. Zhang, JHEP \textbf{03}, 009 (2011).
\newblock \doi{10.1007/JHEP03(2011)009}

\bibitem{Kandemir:2016pjf}
B.S. Kandemir, U.~Ertem, Annalen Phys. \textbf{529}, 1600330 (2017).
\newblock \doi{10.1002/andp.201600330}

\bibitem{Panotopoulos:2018can}
G.~Panotopoulos, Gen. Rel. Grav. \textbf{50}, 59 (2018).
\newblock \doi{10.1007/s10714-018-2381-5}

\bibitem{Rincon:2018sgd}
A.~Rinc\'on, G.~Panotopoulos, Phys. Rev. D \textbf{97}, 024027 (2018).
\newblock \doi{10.1103/PhysRevD.97.024027}

\bibitem{Dias:2019ery}
O.J.C.~Dias, H.S.~Reall and J.E.~Santos,
JHEP \textbf{12}, 097 (2019).
\newblock \doi{10.1007/JHEP12(2019)097} 

\bibitem{Singha:2022bvr}
C.~Singha, S.~Chakraborty, N.~Dadhich, JHEP \textbf{06}, 028 (2022).
\newblock \doi{10.1007/JHEP06(2022)028}

\bibitem{Katagiri:2022qje}
T.~Katagiri, M.~Kimura, Phys. Rev. D \textbf{106}, 044052 (2022).
\newblock \doi{10.1103/PhysRevD.106.044052}

\bibitem{DalBoscoFontana:2023syy}
R.~Dal Bosco Fontana,
Class. Quant. Grav. \textbf{41}, 145010 (2024). 
\newblock \doi{10.1088/1361-6382/ad5782}

\bibitem{Fontana:2023dix}
R.D.B. Fontana,  Phys. Rev. D {\textbf{109}}, 044039   (2024).
\newblock \doi{10.1103/PhysRevD.109.044039}

\bibitem{Gonzalez:2021vwp}
P.A. Gonz\'alez, A.~Rinc\'on, J.~Saavedra, Y.~V\'asquez, Phys. Rev. D
  \textbf{104}, 084047 (2021).
\newblock \doi{10.1103/PhysRevD.104.084047}

\bibitem{Steif:1993zv}
A.R. Steif, Phys. Rev. D \textbf{49}, 585 (1994).
\newblock \doi{10.1103/PhysRevD.49.R585}

\bibitem{Lifschytz:1993eb}
G.~Lifschytz, M.~Ortiz, Phys. Rev. D \textbf{49}, 1929 (1994).
\newblock \doi{10.1103/PhysRevD.49.1929}

\bibitem{Kothawala:2008sm}
D.A. Kothawala, S.~Shankaranarayanan, L.~Sriramkumar, JHEP \textbf{09}, 095
  (2008).
\newblock \doi{10.1088/1126-6708/2008/09/095}

\bibitem{Shiraishi:1993nu}
K.~Shiraishi, T.~Maki, Class. Quant. Grav. \textbf{11}, 695 (1994).
\newblock \doi{10.1088/0264-9381/11/3/019}

\bibitem{Shiraishi:1993qnr}
K.~Shiraishi, T.~Maki, Phys. Rev. D \textbf{49}, 5286 (1994).
\newblock \doi{10.1103/PhysRevD.49.5286}

\bibitem{Barroso:2019cwp}
V.S. Barroso, J.P.M. Pitelli, Gen. Rel. Grav. \textbf{52}, 29 (2020).
\newblock \doi{10.1007/s10714-020-02672-4}

\bibitem{Morley:2020ayr}
T.~Morley, P.~Taylor, E.~Winstanley, Class. Quant. Grav. \textbf{38}, 035009
  (2021).
\newblock \doi{10.1088/1361-6382/aba58a}

\bibitem{Namasivayam:2022bky}
S.~Namasivayam, E.~Winstanley, Gen. Rel. Grav. \textbf{55}, 13 (2023).
\newblock \doi{10.1007/s10714-022-03056-6}

\bibitem{Frolov:1989jh}
V.P. Frolov, K.S. Thorne, Phys. Rev. D \textbf{39}, 2125 (1989).
\newblock \doi{10.1103/PhysRevD.39.2125}

\bibitem{Ottewill:2000qh}
A.C. Ottewill, E.~Winstanley, Phys. Rev. D \textbf{62}, 084018 (2000).
\newblock \doi{10.1103/PhysRevD.62.084018}

\bibitem{Duffy:2005mz}
G.~Duffy, A.C. Ottewill, Phys. Rev. D \textbf{77}, 024007 (2008).
\newblock \doi{10.1103/PhysRevD.77.024007}

\bibitem{Casals:2005kr}
M.~Casals, A.C. Ottewill, Phys. Rev. D \textbf{71}, 124016 (2005).
\newblock \doi{10.1103/PhysRevD.71.124016}

\bibitem{Casals:2012es}
M.~Casals, S.R. Dolan, B.C. Nolan, A.C. Ottewill, E.~Winstanley, Phys. Rev. D
  \textbf{87}, 064027 (2013).
\newblock \doi{10.1103/PhysRevD.87.064027}

\bibitem{Balakumar:2022yvx}
V.~Balakumar, R.P. Bernar, E.~Winstanley, Phys. Rev. D \textbf{106}, 125013
  (2022).
\newblock \doi{10.1103/PhysRevD.106.125013}


\end{thebibliography}

\end{document}